\pdfoutput=1
\documentclass[a4paper,american,floatfix,pdftex,superscriptaddress,twopage,%
aip,jcp,
reprint,final,twocolumn%
]{revtex4-1}%
\usepackage{amsmath,amssymb}
\usepackage{babel}
\usepackage{graphicx}
\usepackage[T1]{fontenc}
\usepackage[utf8]{inputenc}
\usepackage[textsize=footnotesize]{todonotes}
\usepackage{hyperref, hypernat}
\usepackage[todos]{CMImacros}

\graphicspath{{./fig/}} % define figure directory

\makeatletter%
\renewcommand*{\@fnsymbol}[1]{\ensuremath{\ifcase#1\or \mathsection\or \|\or \mathparagraph\or *\or
      **\or \dagger\or \ddagger\or \dagger\dagger \or \ddagger\ddagger \else\@ctrerr\fi}}
\makeatother%

\newcommand*{\cisMVK}{\emph{s-cis}\ MVK\xspace}
\newcommand*{\transMVK}{\emph{s-trans}\ MVK\xspace}
\newcommand*{\suppinf}{supplementary information\xspace}%
\newcommand*{\suppfig}[1]{Figure~S#1 in the supplementary information\xspace}%

\newcommand{\cfeldesy}{\affiliation{Center for Free-Electron Laser Science, Deutsches
      Elektronen-Synchrotron DESY, Notkestrasse 85, 22607 Hamburg, Germany}}%
\newcommand*{\basel}{\affiliation{Department of Chemistry, University of Basel, Klingelbergstrasse
      80, 4056 Basel, Switzerland}}%
\newcommand{\uhhcui}{\affiliation{Center for Ultrafast Imaging, Universität Hamburg, Luruper
      Chaussee 149, 22761 Hamburg, Germany}}%
\newcommand{\uhhphys}{\affiliation{Department of Physics, Universität Hamburg, Luruper Chaussee 149,
      22761 Hamburg, Germany}}%
\newcommand*{\tsinghua}{\altaffiliation[Permanent address: ]{Department of Physics, Tsinghua
      University, 100084, Beijing, China}}%
\newcommand*{\jilinu}{\altaffiliation[Permanent address: ]{Institute of Atomic and Molecular
      Physics, Jilin University, Changchun 130012, China}}%%
\newcommand*{\jkemail}{\email[Corresponding author. Email:~]{jochen.kuepper@cfel.de}}%
\newcommand*{\swemail}{\email[Email:~]{stefan.willitsch@unibas.ch}}%
\newcommand*{\cmiweb}{\homepage[website:~]{https://www.controlled-molecule-imaging.org}}%

\begin{document}
\title{Spatial Separation of the Conformers of Methyl Vinyl Ketone}%
\author{Jia Wang}\tsinghua\cfeldesy%
\author{Ardita Kilaj}\basel%
\author{Lanhai He}\jilinu\cfeldesy%
\author{Karol Długołęcki}\cfeldesy%
\author{Stefan Willitsch}\swemail\basel%
\author{Jochen Küpper}\jkemail\cmiweb\cfeldesy\uhhphys\uhhcui%
%\date{\today}%
\begin{abstract}\noindent%
   Methyl vinyl ketone (C$_4$H$_6$O) is a volatile, labile organic compound of importance in
   atmospheric chemistry. We prepared a molecular beam of methyl vinyl ketone with a rotational
   temperature of 1.2(2)~K and demonstrated the spatial separation of the \emph{s-cis} and
   \emph{s-trans} conformers of methyl vinyl ketone using the electrostatic deflector. The resulting
   sample density was $1.5(2)\times10^{8}~\text{cm}^{-3}$ for the direct beam in the laser
   ionization region. These conformer-selected methyl vinyl ketone samples are well suited for
   conformer-specific chemical reactivity studies such as in Diels-Alder cycloaddition reactions.
\end{abstract}
\maketitle

\section{Introduction}
Methyl vinyl ketone (MVK, 3-butene-2-one, C$_4$H$_6$O) is the simplest $\alpha,\beta$-unsaturated
ketone and an important oxygenated volatile organic compound. MVK results from many sources such as
vehicle exhaust \cite{Biesenthal:GRL24:1375, Zhang:CPL358:171}, biomass
burning~\cite{Brilli:AE97:54} and the ozonolysis of isoprene~\cite{Gutbrod:JACS119:7330}. As a
primary first-yield product of isoprene oxidation in earth's atmosphere~\cite{Tuazon:IJCK21:1141,
   Galloway:ACP11:10779}, MVK remains in the gas phase and is highly
reactive~\cite{Pierotti:JGRA95:1871}. As a result, MVK has an important impact on the photochemical
activity in the boundary layer, in particular in forested areas~\cite{Karl:JAC55:167}, and
contributes to the destruction of ozone due to its formation of species such as formaldehyde and
methylglyoxal~\cite{Gutbrod:JACS119:7330, Praske:JPCA119:4562}. The atmospheric lifetime of MVK is
$\ordsim10$~h~\cite{Karl:JAC55:167, Atkinson:ACP6:3625} due to its fast reaction with hydroxyl
radicals under atmospheric conditions~\cite{Atkinson:CR86:69}. On the other hand, in the troposphere
isoprene reacts with hydroxyl radicals and ozone molecules leading to a significant yield of
MVK~\cite{Aschmann:EST28:1539}, which is important for the formation of secondary organic aerosols
and the overall NO$_{x}$ cycle~\cite{Kroll:AE42:3593, Praske:JPCA119:4562}. Moreover, MVK is a
candidate for prototypical pericyclic reactions, such as the Diels-Alder (DA)
cycloaddition~\cite{Jorgensen:JACS115:2936, Jorgensen:JCSFT90:1727}.

The ultraviolet absorption~\cite{Rogers:JACS69:2544}, microwave~\cite{Foster:JCP43:1064,
   Fantoni:CPL133:27, DeSmedt:JMS195:227} and infrared~\cite{Noack:CJC39:2225, Bowles:JCSB:810,
   Durig:JCP75:3660, Oelichmann:JMS77:179} spectra of MVK provided evidence for a mixture of
\emph{s-cis} and \emph{s-trans} MVK conformers and showed that \transMVK is more stable than
\cisMVK. Recently, the high-resolution rotational (7.5--18.5~GHz)~\cite{Wilcox:CPL508:10},
millimeter-wave~\cite{Zakharenko:JPCA121:6420}, and infrared spectra
(540--6500~\invcm)~\cite{Lindenmaier:JPCA121:1195} of MVK were reported. By combining experimental
data and high-level quantum-chemistry calculations, the relative energy of \transMVK and \cisMVK was
determined as $164\pm30$~\invcm~\cite{Zakharenko:JPCA121:6420}, yielding an equilibrium mixture of
approximately 69~\% \transMVK and 31~\% \cisMVK at room temperature~\cite{Zakharenko:JPCA121:6420,
   Lindenmaier:JPCA121:1195}.

The reactivity of different conformers may vary significantly~\cite{Chang:Science342:98,
   Willitsch:ACP162:307}. Neutral molecules can be manipulated in the gas-phase using the
electrostatic deflection technique~\cite{Chang:IRPC34:557}, which was demonstrated for the
separation of individual quantum states~\cite{Nielsen:PCCP13:18971, Horke:ACIE53:11965,
   Kienitz:JCP147:024304}, conformers~\cite{Filsinger:PRL100:133003, Filsinger:ACIE48:6900,
   Kierspel:CPL591:130, Teschmit:ACIE57:13775, You:JPCA122:1194}, or molecular
clusters~\cite{Trippel:PRA86:033202, You:JPCA122:1194, Johny:CPL721:149}. This separation can be
exploited for the investigation of the specific chemical reactivities of individual molecular
species~\cite{Chang:Science342:98, Kilaj:NatComm9:2096, Roesch:JCP140:124202, Kilaj:PCCP:inprep}.

Here, we demonstrate the preparation of a cold and dense molecular beam of MVK and the spatial
separation of the \emph{s-cis} and \emph{s-trans} conformers using the electrostatic deflector.
Furthermore, we determine the density of the produced cold samples. The spatially separated
conformers of MVK could be used for non-species-specific experiments, \eg, conformer-specific
reactivity studies~\cite{Chang:Science342:98, Kilaj:NatComm9:2096} or ultrafast structural imaging
experiments~\cite{Kuepper:PRL112:083002, Hensley:PRL109:133202, Wiese:PRR2020:inprep}.

\section{Experimental Methods}
\label{sec:setup}
The experimental setup was described previously~\cite{Teschmit:ACIE57:13775, Wang:JMS1208:127863}. A
homebuilt gas handing system with a rotating high-pressure cylinder was added to fully and
permanently mix MVK (Sigma-Aldrich, 99~\%, used without further purification) and helium, see
\suppfig{1}. The reservoir was filled with 2~ml of MVK, de-aired down to $\ordsim10^{-2}$~mbar, and
the MVK vapor was mixed with helium gas at 20~bar. The gas mixture was supersonically expanded
through a cantilever piezo valve~\cite{Irimia:RSI80:113303} operated at a repetition rate of 20 Hz.
Helium seed gas was used for best rotational cooling and minimal longitudinal dispersion of the beam
and correspondingly high beam density in the interaction region for the planned conformer-specific
chemical-reaction experiments. Two skimmers, placed 55~mm ($\varnothing=3$~mm) and 365~mm
($\varnothing=1.5$~mm) downstream of the valve were used to collimate the molecular beam, which was
then directed through the electrostatic deflector~\cite{Chang:IRPC34:557} before passing through a
third skimmer ($\varnothing=1.5$~mm) 562~mm downstream of the nozzle. MVK was ionized by a
femtosecond laser with a wavelength centered at $\ordsim800$~nm and a pulse duration of 45~fs
(full-width at half maximum, FWHM) that was focused to $44(4)~\um$ (FWHM, $\omega_0=37(3)~\um$) in
the interaction region by a $f=500$~mm lens. The resulting ions were detected by a two-plate
time-of-flight (TOF) mass spectrometer (MS)~\cite{Kienitz:CPC17:3740}.

\section{Results and Discussion}
\label{sec:results}
\begin{figure}[b]
   \includegraphics[width=\linewidth]{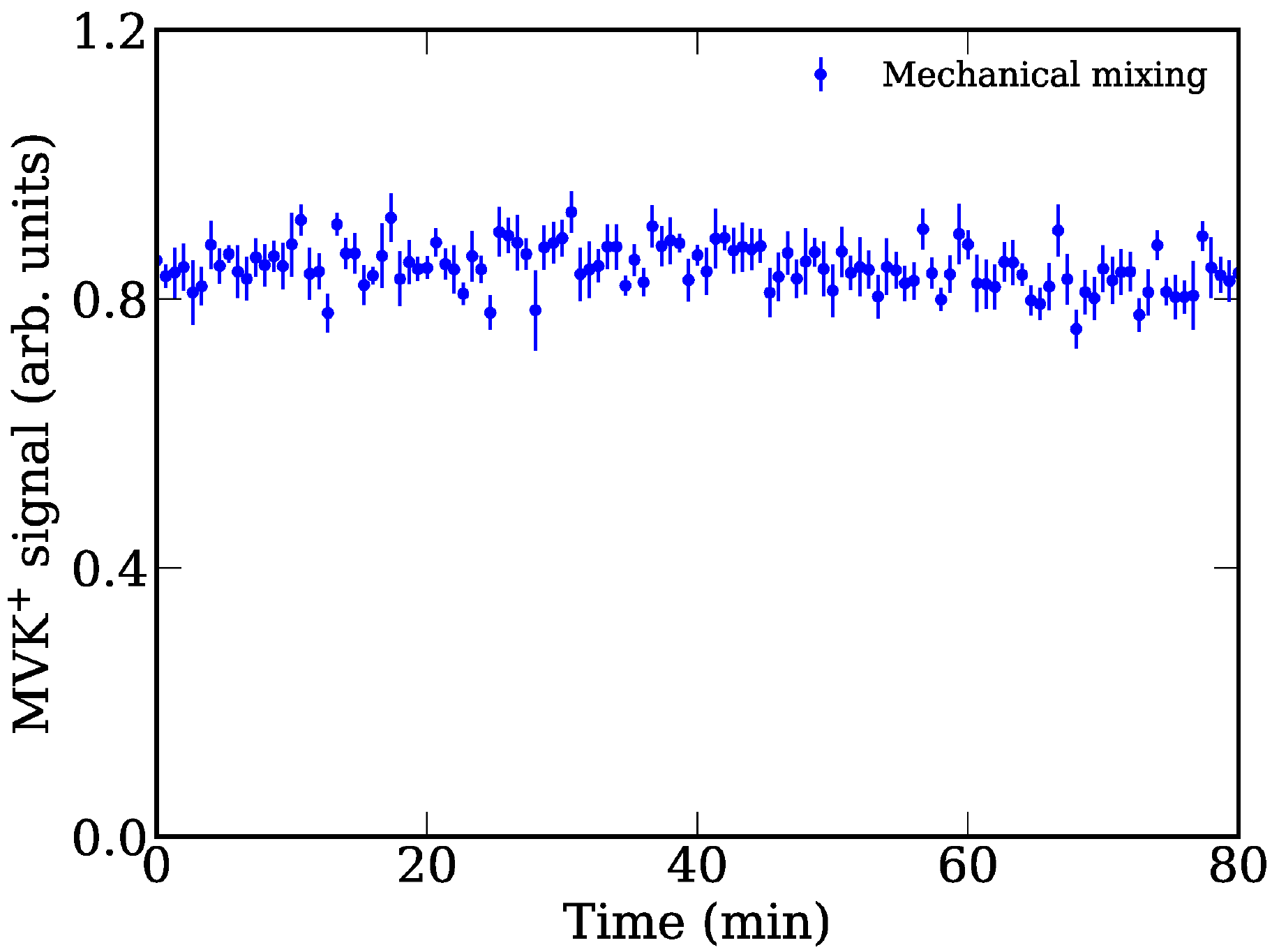}
   \caption{MVK signal over time for the sample preparation (200~ppm) of mechanical mixing by
      rotation of the sample cylinder.}
   \label{fig:mvkstable}
\end{figure}
MVK is a liquid at room temperature and condenses on the sample reservoir walls, resulting in fast
demixing of the prepared gas mixtures and corresponding fast decays of the MVK density in the
molecular beam. Demixing was avoided through rotation of the sample reservoir (\emph{vide supra})
and \autoref{fig:mvkstable} shows the resulting stability of the MVK signal following 20~h of sample
rotation. Under these conditions the sample density was stable over a few hours and it decreased to
70~\% over four days.

The normalized experimental vertical molecular beam profiles of MVK for different backing pressures
of 2, 4, 6, and 8~bar are shown in \suppfig{5}. The full width of the direct molecular beam (0~kV)
is 2~mm, determined by the skimmers and the distance between the third skimmer and ionization point.
The deflected beam (10~kV) is deflected upward by $\ordsim0.8$~mm when a pressure of 2~bar is
applied to the piezo valve. Increasing the pressure to 4~bar, 6~bar, and 8~bar the deflection of the
beam increases to $\ordsim1.0$~mm, $\ordsim1.1$~mm, and $\ordsim1.1$~mm, respectively, which is due
to the correspondingly lower rotational temperature of these beams~\cite{Filsinger:JCP131:064309,
   Chang:IRPC34:557}.

\begin{figure}
   \includegraphics[width=\linewidth]{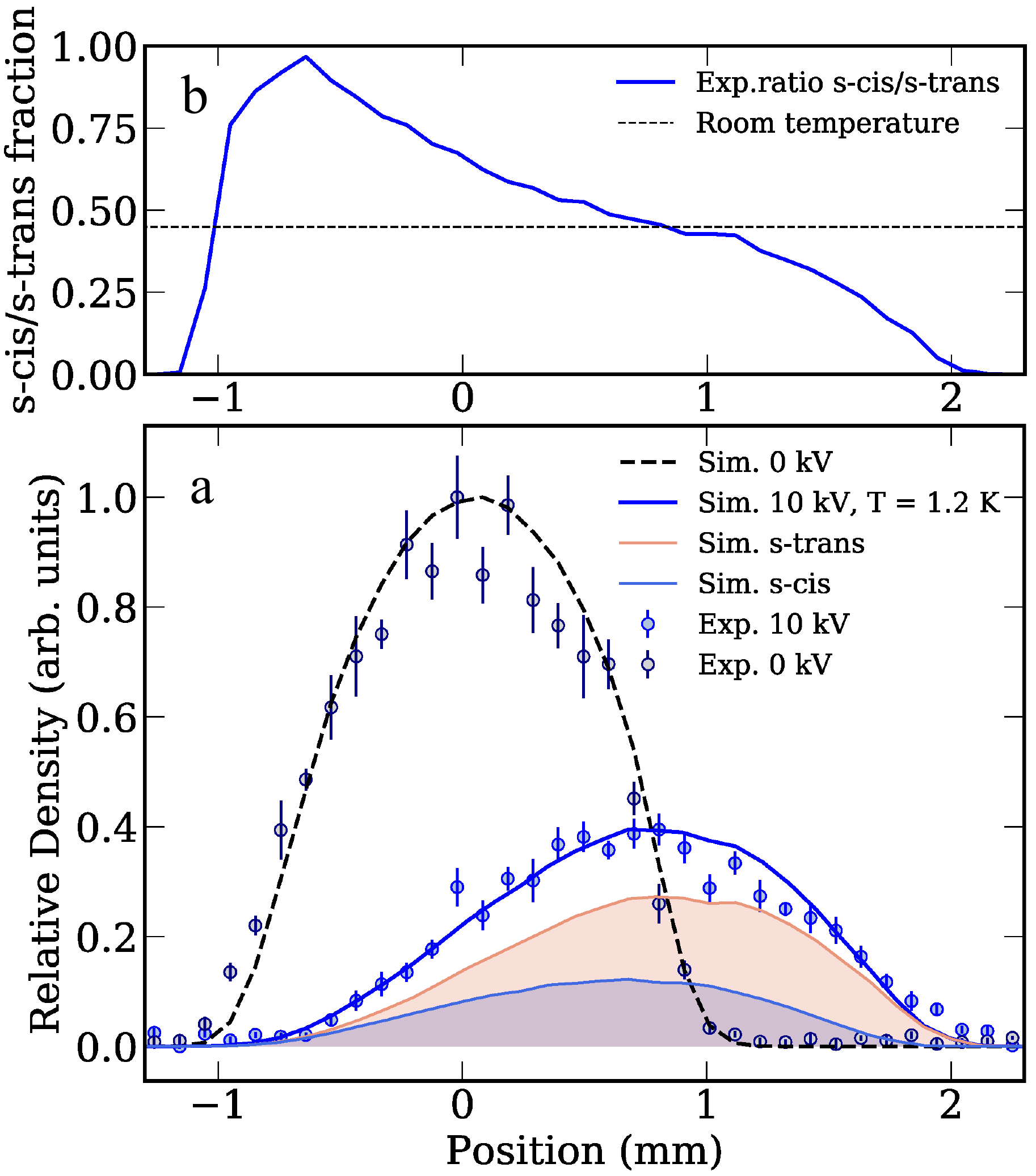}%
   \caption{(a) Direct and deflected spatial beam profiles of MVK and simulation results. (b) The
      fractional population of the \emph{s-cis} and \emph{s-trans} conformer in the beam; the thin
      gray line depicts the ratio at room temperature~[\onlinecite{Zakharenko:JPCA121:6420,
         Lindenmaier:JPCA121:1195}].}
   \label{fig:deflection}
\end{figure}
\autoref[a]{fig:deflection} shows the experimental and simulated molecular beam profiles of MVK
seeded in 8~bar of helium for deflector voltages of 0~V and 10~kV. The mass spectrum of the direct
(0~kV) and deflected (10~kV) molecular beam are shown in \suppfig{4}. The spectrum of the deflected
beam mainly contains signals from the MVK parent ion M$^{+}$ ($m/z=70$) and from fragment ions
[M-CH$_{2}$$=$CH]$^{+}$ ($m/z=43$) and [M-CH$_{3}$]$^{+}$ ($m/z=55$).

Solid and dotted lines in \autoref[a]{fig:deflection} show simulated spatial profiles using the
molecular parameters and calculations detailed in the \suppinf. The effective dipole moments of the
rotational states of \emph{s-trans} are larger than ones of \emph{s-cis} and, therefore,
\emph{s-trans} deflects further than \emph{s-cis}. Assuming a thermal distribution of the population
of rotational states, the best fit for the profile of MVK in \autoref{fig:deflection} was obtained
for a rotational temperature of 1.2(2)~K. The deflection and simulation profiles of MVK seeded in
different pressures of helium are shown in \suppfig{3}.

Although no full separation was possible, \emph{s-trans} MVK was deflected more than \emph{s-cis}
MVK. The fractional contributions of the conformers across the vertical beam profile are shown in
\autoref[b]{fig:deflection}, assuming the same excitation and ionization cross-sections for the two
conformers~\cite{Kierspel:CPL591:130}. A beam of \emph{s-trans} conformer with a purity higher than
90~\% was obtained for vertical molecular-beam positions $y\geq1.9$~mm.

\begin{figure}
   \includegraphics[width=\linewidth]{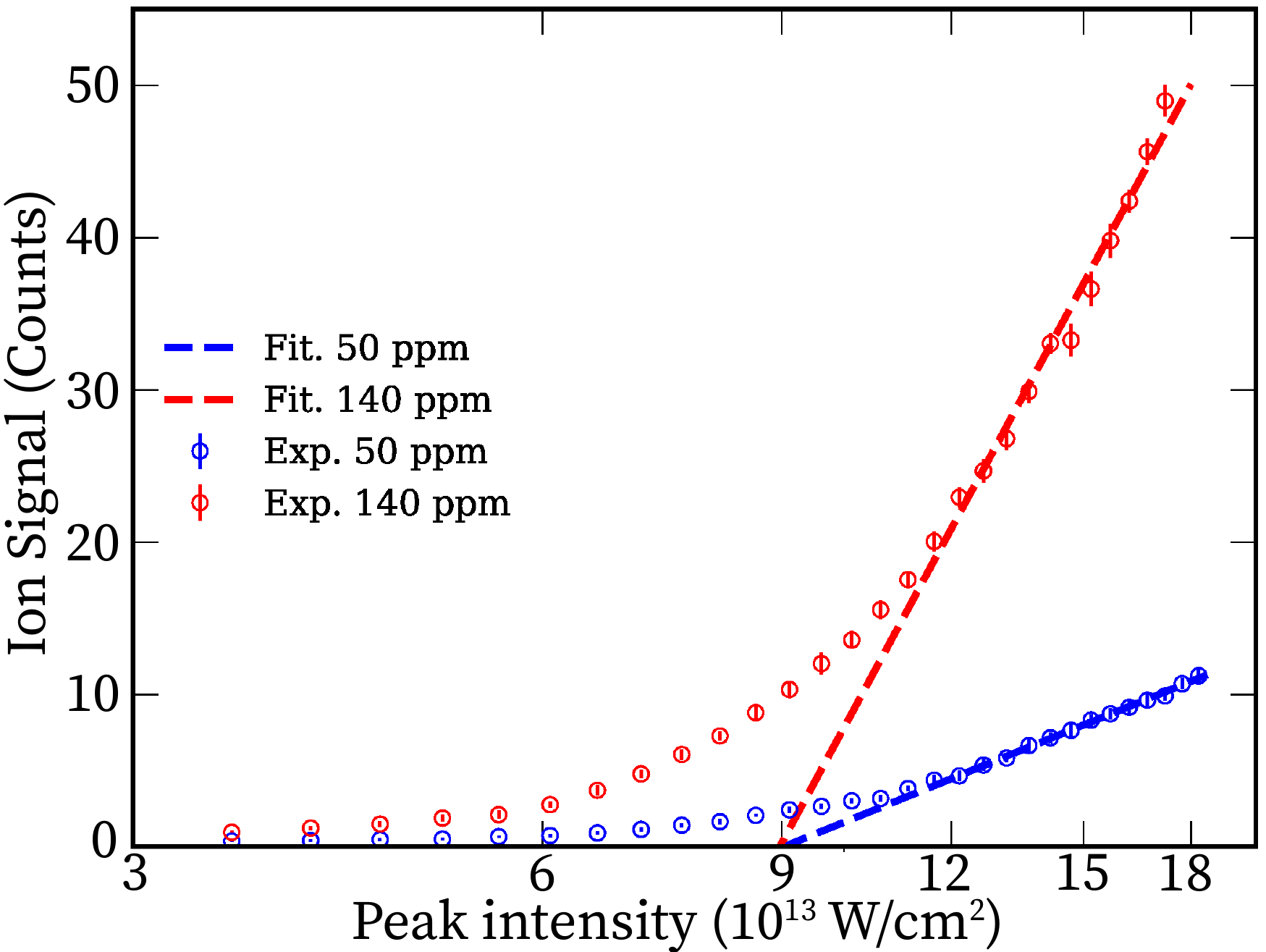}
   \caption{MVK signal, in ion counts per laser shot, of the direct molecular beam at vertical
      molecular-beam position 0~mm versus the peak intensity on a semi-logarithmic scale (dots). MVK
      was seeded in 2~bar of He at 50~ppm (blue) and 140~ppm (red). The dashed lines are the result
      of a linear fit to the asymptotic behavior.}
   \label{fig:density}
\end{figure}
The MVK sample densities in the experiments were estimated based on a strong-field ionization
model~\cite{Hankin:PRA64:013405, Wiese:NJP21:083011}. Assuming an instrument sensitivity of 50~\%
for the MCP detector~\cite{Fehre:RSI89:045112} the asymptotic slope of an integral ionization signal
with respect to the natural logarithm of peak intensity can be expressed as
$S=\varrho\pi\sigma^2D$~\cite{Hankin:PRA64:013405, Wiese:NJP21:083011}, where $\varrho$ represents
the sample density, $\sigma$ the standard deviation of the transverse laser intensity distribution
and $D$ the length of the focal volume in the molecular beam. The MVK signal count of the direct
molecular beam at vertical molecular-beam position 0~mm \emph{versus} the peak intensity are shown
on a semi-logarithmic scale in \autoref{fig:density}, where the sum of the signals of the parent ion
($m/z=70$) and its fragments ($m/z=43$ and $m/z=55$) was used. The saturation onset was deduced for
a laser peak intensity $I_\text{sat}=9.0(1)\times10^{13}~\Wpcmcm$ at relative concentrations of 50
and 140~ppm of MVK in He. For the known transverse waist of the laser focus of
$\omega_0=2\sigma=37(3)~\um$ and a
molecular-beam diameter of $D=2.0(3)$~mm the sample densities in the laser ionization region,
$\ordsim750$~mm downstream the valve, were estimated to be $3.3(5)\times10^{7}~\text{cm}^{-3}$ and
$1.5(2)\times10^{8}~\text{cm}^{-3}$, respectively.

From these combined-conformer sample densities, the beam densities of \emph{s-cis} MVK at $-0.7$~mm
and \emph{s-trans} MVK at $+1.9$~mm were estimated to be $\ordsim1.7(3)\times10^{6}~\text{cm}^{-3}$
and $\ordsim4.9(8)\times10^{6}~\text{cm}^{-3}$, respectively. However, the individual conformers'
reaction-rate constants $k_{cis}$ and $k_{trans}$ can be derived from a few reactivity measurements
for intense parts of the beam with different conformer compositions, combined with the known
relative populations from the deflection profiles~\cite{Chang:Science342:98, Kilaj:NatComm9:2096}.

\section{Conclusion}
We demonstrated the use of a rotating-sample reservoir for the production of a dense and cold
molecular beam of MVK with stable molecular densities over more than a day and a rotational
temperature of 1.2(2)~K. This allowed for the spatial dispersion and partial separation of the
\emph{s-cis} and \emph{s-trans} conformers of MVK using the electrostatic deflector. The achieved
direct-beam density in the detection zone was experimentally determined to be
$1.5(2)\times10^{8}~\text{cm}^{-3}$.

We plan to exploit these conformer-selected MVK samples for reactivity studies, \eg, with the ions
of MVK (self-reaction), methyl vinyl ether, and further dienophile cations, to investigate the
mechanism and conformational specificities of ionic Diels-Alder (DA) cycloaddition
reactions~\cite{Rivero:CPL683:598}. Furthermore, the reactivity of the conformers in reactions with
neutral atmospheric molecules, such as OH, would be extremely interesting for atmospheric chemistry
applications.

\bigskip % typical revtex problem

\section*{Supplementary Information}%
See the supplementary material for a schematic of the gas panel, the direct and deflected molecular
beam profiles of MVK for different backing pressures, the mass spectra obtained in the direct and
the deflected molecular beams, the rotational constants and dipole moments of the MVK conformers,
their Stark energies, and the deflection and simulation profiles of MVK seeded in helium of
different pressures.

\begin{acknowledgments}\noindent%
   We acknowledge help with the experimental setup by Jovana Petrovič and Nicolai Pohlmann.

   This work has been supported by the cluster of excellence ``Advanced Imaging of Matter'' (AIM,
   EXC~2056, ID~390715994) of the Deutsche Forschungsgemeinschaft (DFG), by the European Research
   Council under the European Union's Seventh Framework Program (FP7/2007-2013) through the
   Consolidator Grant COMOTION (614507), and by the Swiss National Science Foundation under grants
   nr.\ BSCGI0\_157874 and IZCOZ0\_189907. J.W.\ and L.H.\ acknowledge fellowships within the
   framework of the Helmholtz-OCPC postdoctoral exchange program.
\end{acknowledgments}

\bibliography{string,cmi}

%aipnum4-2.bst 2019-01-14 (MD) hand-edited version of apsrev4-1.bst
\begin{thebibliography}{53}%
\makeatletter
\providecommand \@ifxundefined [1]{%
 \@ifx{#1\undefined}
}%
\providecommand \@ifnum [1]{%
 \ifnum #1\expandafter \@firstoftwo
 \else \expandafter \@secondoftwo
 \fi
}%
\providecommand \@ifx [1]{%
 \ifx #1\expandafter \@firstoftwo
 \else \expandafter \@secondoftwo
 \fi
}%
\providecommand \natexlab [1]{#1}%
\providecommand \enquote  [1]{``#1''}%
\providecommand \bibnamefont  [1]{#1}%
\providecommand \bibfnamefont [1]{#1}%
\providecommand \citenamefont [1]{#1}%
\providecommand \href@noop [0]{\@secondoftwo}%
\providecommand \href [0]{\begingroup \@sanitize@url \@href}%
\providecommand \@href[1]{\@@startlink{#1}\@@href}%
\providecommand \@@href[1]{\endgroup#1\@@endlink}%
\providecommand \@sanitize@url [0]{\catcode `\\12\catcode `\$12\catcode
  `\&12\catcode `\#12\catcode `\^12\catcode `\_12\catcode `\%12\relax}%
\providecommand \@@startlink[1]{}%
\providecommand \@@endlink[0]{}%
\providecommand \url  [0]{\begingroup\@sanitize@url \@url }%
\providecommand \@url [1]{\endgroup\@href {#1}{\urlprefix }}%
\providecommand \urlprefix  [0]{URL }%
\providecommand \Eprint [0]{\href }%
\providecommand \doibase [0]{https://doi.org/}%
\providecommand \selectlanguage [0]{\@gobble}%
\providecommand \bibinfo  [0]{\@secondoftwo}%
\providecommand \bibfield  [0]{\@secondoftwo}%
\providecommand \translation [1]{[#1]}%
\providecommand \BibitemOpen [0]{}%
\providecommand \bibitemStop [0]{}%
\providecommand \bibitemNoStop [0]{.\EOS\space}%
\providecommand \EOS [0]{\spacefactor3000\relax}%
\providecommand \BibitemShut  [1]{\csname bibitem#1\endcsname}%
\let\auto@bib@innerbib\@empty
%</preamble>
\bibitem [{\citenamefont {Biesenthal}\ and\ \citenamefont
  {Shepson}(1997)}]{Biesenthal:GRL24:1375}%
  \BibitemOpen
  \bibfield  {author} {\bibinfo {author} {\bibfnamefont {T.~A.}\ \bibnamefont
  {Biesenthal}}\ and\ \bibinfo {author} {\bibfnamefont {P.~B.}\ \bibnamefont
  {Shepson}},\ }\bibfield  {title} {\enquote {\bibinfo {title} {Observations of
  anthropogenic inputs of the isoprene oxidation products methyl vinyl ketone
  and methacrolein to the atmosphere},}\ }\href
  {https://doi.org/10.1029/97GL01337} {\bibfield  {journal} {\bibinfo
  {journal} {Geophys. Res. Lett.}\ }\textbf {\bibinfo {volume} {24}},\ \bibinfo
  {pages} {1375--1378} (\bibinfo {year} {1997})}\BibitemShut {NoStop}%
\bibitem [{\citenamefont {Zhang}, \citenamefont {Lei},\ and\ \citenamefont
  {Zhang}(2002)}]{Zhang:CPL358:171}%
  \BibitemOpen
  \bibfield  {author} {\bibinfo {author} {\bibfnamefont {D.}~\bibnamefont
  {Zhang}}, \bibinfo {author} {\bibfnamefont {W.}~\bibnamefont {Lei}},\ and\
  \bibinfo {author} {\bibfnamefont {R.}~\bibnamefont {Zhang}},\ }\bibfield
  {title} {\enquote {\bibinfo {title} {Mechanism of {OH} formation from
  ozonolysis of isoprene: kinetics and product yields},}\ }\href
  {https://doi.org/https://doi.org/10.1016/S0009-2614(02)00260-9} {\bibfield
  {journal} {\bibinfo  {journal} {Chem. Phys. Lett.}\ }\textbf {\bibinfo
  {volume} {358}},\ \bibinfo {pages} {171 -- 179} (\bibinfo {year}
  {2002})}\BibitemShut {NoStop}%
\bibitem [{\citenamefont {Brilli}\ \emph {et~al.}(2014)\citenamefont {Brilli},
  \citenamefont {Gioli}, \citenamefont {Ciccioli}, \citenamefont {Zona},
  \citenamefont {Loreto}, \citenamefont {Janssens},\ and\ \citenamefont
  {Ceulemans}}]{Brilli:AE97:54}%
  \BibitemOpen
  \bibfield  {author} {\bibinfo {author} {\bibfnamefont {F.}~\bibnamefont
  {Brilli}}, \bibinfo {author} {\bibfnamefont {B.}~\bibnamefont {Gioli}},
  \bibinfo {author} {\bibfnamefont {P.}~\bibnamefont {Ciccioli}}, \bibinfo
  {author} {\bibfnamefont {D.}~\bibnamefont {Zona}}, \bibinfo {author}
  {\bibfnamefont {F.}~\bibnamefont {Loreto}}, \bibinfo {author} {\bibfnamefont
  {I.~A.}\ \bibnamefont {Janssens}},\ and\ \bibinfo {author} {\bibfnamefont
  {R.}~\bibnamefont {Ceulemans}},\ }\bibfield  {title} {\enquote {\bibinfo
  {title} {Proton transfer reaction time-of-flight mass spectrometric
  ({PTR-TOF-MS}) determination of volatile organic compounds ({VOC}s) emitted
  from a biomass fire developed under stable nocturnal conditions},}\ }\href
  {https://doi.org/https://doi.org/10.1016/j.atmosenv.2014.08.007} {\bibfield
  {journal} {\bibinfo  {journal} {Atmos. Environ.}\ }\textbf {\bibinfo {volume}
  {97}},\ \bibinfo {pages} {54 -- 67} (\bibinfo {year} {2014})}\BibitemShut
  {NoStop}%
\bibitem [{\citenamefont {Gutbrod}\ \emph {et~al.}(1997)\citenamefont
  {Gutbrod}, \citenamefont {Kraka}, \citenamefont {Schindler},\ and\
  \citenamefont {Cremer}}]{Gutbrod:JACS119:7330}%
  \BibitemOpen
  \bibfield  {author} {\bibinfo {author} {\bibfnamefont {R.}~\bibnamefont
  {Gutbrod}}, \bibinfo {author} {\bibfnamefont {E.}~\bibnamefont {Kraka}},
  \bibinfo {author} {\bibfnamefont {R.~N.}\ \bibnamefont {Schindler}},\ and\
  \bibinfo {author} {\bibfnamefont {D.}~\bibnamefont {Cremer}},\ }\bibfield
  {title} {\enquote {\bibinfo {title} {Kinetic and theoretical investigation of
  the gas-phase ozonolysis of isoprene: Carbonyl oxides as an important source
  for {OH} radicals in the atmosphere},}\ }\href
  {https://doi.org/10.1021/ja970050c} {\bibfield  {journal} {\bibinfo
  {journal} {J. Am. Chem. Soc.}\ }\textbf {\bibinfo {volume} {119}},\ \bibinfo
  {pages} {7330--7342} (\bibinfo {year} {1997})}\BibitemShut {NoStop}%
\bibitem [{\citenamefont {Tuazon}\ and\ \citenamefont
  {Atkinson}(1989)}]{Tuazon:IJCK21:1141}%
  \BibitemOpen
  \bibfield  {author} {\bibinfo {author} {\bibfnamefont {E.~C.}\ \bibnamefont
  {Tuazon}}\ and\ \bibinfo {author} {\bibfnamefont {R.}~\bibnamefont
  {Atkinson}},\ }\bibfield  {title} {\enquote {\bibinfo {title} {A product
  study of the gas-phase reaction of methyl vinyl ketone with the {OH} radical
  in the presence of {NO}$_{x}$},}\ }\href
  {https://doi.org/10.1002/kin.550211207} {\bibfield  {journal} {\bibinfo
  {journal} {Int. J. Chem. Kinet.}\ }\textbf {\bibinfo {volume} {21}},\
  \bibinfo {pages} {1141--1152} (\bibinfo {year} {1989})}\BibitemShut {NoStop}%
\bibitem [{\citenamefont {Galloway}\ \emph {et~al.}(2011)\citenamefont
  {Galloway}, \citenamefont {Huisman}, \citenamefont {Yee}, \citenamefont
  {Chan}, \citenamefont {Loza}, \citenamefont {Seinfeld},\ and\ \citenamefont
  {Keutsch}}]{Galloway:ACP11:10779}%
  \BibitemOpen
  \bibfield  {author} {\bibinfo {author} {\bibfnamefont {M.~M.}\ \bibnamefont
  {Galloway}}, \bibinfo {author} {\bibfnamefont {A.~J.}\ \bibnamefont
  {Huisman}}, \bibinfo {author} {\bibfnamefont {L.~D.}\ \bibnamefont {Yee}},
  \bibinfo {author} {\bibfnamefont {A.~W.~H.}\ \bibnamefont {Chan}}, \bibinfo
  {author} {\bibfnamefont {C.~L.}\ \bibnamefont {Loza}}, \bibinfo {author}
  {\bibfnamefont {J.~H.}\ \bibnamefont {Seinfeld}},\ and\ \bibinfo {author}
  {\bibfnamefont {F.~N.}\ \bibnamefont {Keutsch}},\ }\bibfield  {title}
  {\enquote {\bibinfo {title} {Yields of oxidized volatile organic compounds
  during the {OH} radical initiated oxidation of isoprene, methyl vinyl ketone,
  and methacrolein under high-{NO}$_{x}$ conditions},}\ }\href
  {https://doi.org/10.5194/acp-11-10779-2011} {\bibfield  {journal} {\bibinfo
  {journal} {Atmos. Chem. Phys.}\ }\textbf {\bibinfo {volume} {11}},\ \bibinfo
  {pages} {10779--10790} (\bibinfo {year} {2011})}\BibitemShut {NoStop}%
\bibitem [{\citenamefont {Pierotti}\ \emph {et~al.}(1990)\citenamefont
  {Pierotti}, \citenamefont {Wofsy}, \citenamefont {Jacob},\ and\ \citenamefont
  {Rasmussen}}]{Pierotti:JGRA95:1871}%
  \BibitemOpen
  \bibfield  {author} {\bibinfo {author} {\bibfnamefont {D.}~\bibnamefont
  {Pierotti}}, \bibinfo {author} {\bibfnamefont {S.~C.}\ \bibnamefont {Wofsy}},
  \bibinfo {author} {\bibfnamefont {D.}~\bibnamefont {Jacob}},\ and\ \bibinfo
  {author} {\bibfnamefont {R.~A.}\ \bibnamefont {Rasmussen}},\ }\bibfield
  {title} {\enquote {\bibinfo {title} {Isoprene and its oxidation products:
  Methacrolein and methyl vinyl ketone},}\ }\href
  {https://doi.org/10.1029/JD095iD02p01871} {\bibfield  {journal} {\bibinfo
  {journal} {J. Geophys. Res. Atmos.}\ }\textbf {\bibinfo {volume} {95}},\
  \bibinfo {pages} {1871--1881} (\bibinfo {year} {1990})}\BibitemShut {NoStop}%
\bibitem [{\citenamefont {Karl}\ \emph {et~al.}(2006)\citenamefont {Karl},
  \citenamefont {Dorn}, \citenamefont {Holland}, \citenamefont {Koppmann},
  \citenamefont {Poppe}, \citenamefont {Rupp}, \citenamefont {Schaub},\ and\
  \citenamefont {Wahner}}]{Karl:JAC55:167}%
  \BibitemOpen
  \bibfield  {author} {\bibinfo {author} {\bibfnamefont {M.}~\bibnamefont
  {Karl}}, \bibinfo {author} {\bibfnamefont {H.-P.}\ \bibnamefont {Dorn}},
  \bibinfo {author} {\bibfnamefont {F.}~\bibnamefont {Holland}}, \bibinfo
  {author} {\bibfnamefont {R.}~\bibnamefont {Koppmann}}, \bibinfo {author}
  {\bibfnamefont {D.}~\bibnamefont {Poppe}}, \bibinfo {author} {\bibfnamefont
  {L.}~\bibnamefont {Rupp}}, \bibinfo {author} {\bibfnamefont {A.}~\bibnamefont
  {Schaub}},\ and\ \bibinfo {author} {\bibfnamefont {A.}~\bibnamefont
  {Wahner}},\ }\bibfield  {title} {\enquote {\bibinfo {title} {Product study of
  the reaction of {OH} radicals with isoprene in the atmosphere simulation
  chamber {SAPHIR}},}\ }\href {https://doi.org/10.1007/s10874-006-9034-x}
  {\bibfield  {journal} {\bibinfo  {journal} {J . Atmos. Chem.}\ }\textbf
  {\bibinfo {volume} {55}},\ \bibinfo {pages} {167--187} (\bibinfo {year}
  {2006})}\BibitemShut {NoStop}%
\bibitem [{\citenamefont {Praske}\ \emph {et~al.}(2015)\citenamefont {Praske},
  \citenamefont {Crounse}, \citenamefont {Bates}, \citenamefont {Kurtén},
  \citenamefont {Kjaergaard},\ and\ \citenamefont
  {Wennberg}}]{Praske:JPCA119:4562}%
  \BibitemOpen
  \bibfield  {author} {\bibinfo {author} {\bibfnamefont {E.}~\bibnamefont
  {Praske}}, \bibinfo {author} {\bibfnamefont {J.~D.}\ \bibnamefont {Crounse}},
  \bibinfo {author} {\bibfnamefont {K.~H.}\ \bibnamefont {Bates}}, \bibinfo
  {author} {\bibfnamefont {T.}~\bibnamefont {Kurtén}}, \bibinfo {author}
  {\bibfnamefont {H.~G.}\ \bibnamefont {Kjaergaard}},\ and\ \bibinfo {author}
  {\bibfnamefont {P.~O.}\ \bibnamefont {Wennberg}},\ }\bibfield  {title}
  {\enquote {\bibinfo {title} {Atmospheric fate of methyl vinyl ketone: Peroxy
  radical reactions with {NO} and {HO}$_{2}$},}\ }\href
  {https://doi.org/10.1021/jp5107058} {\bibfield  {journal} {\bibinfo
  {journal} {J. Phys. Chem. A}\ }\textbf {\bibinfo {volume} {119}},\ \bibinfo
  {pages} {4562--4572} (\bibinfo {year} {2015})},\ \bibinfo {note} {pMID:
  25486386}\BibitemShut {NoStop}%
\bibitem [{\citenamefont {Atkinson}\ \emph {et~al.}(2006)\citenamefont
  {Atkinson}, \citenamefont {Baulch}, \citenamefont {Cox}, \citenamefont
  {Crowley}, \citenamefont {Hampson}, \citenamefont {Hynes}, \citenamefont
  {Jenkin}, \citenamefont {Rossi}, \citenamefont {Troe},\ and\ \citenamefont
  {Subcommittee}}]{Atkinson:ACP6:3625}%
  \BibitemOpen
  \bibfield  {author} {\bibinfo {author} {\bibfnamefont {R.}~\bibnamefont
  {Atkinson}}, \bibinfo {author} {\bibfnamefont {D.~L.}\ \bibnamefont
  {Baulch}}, \bibinfo {author} {\bibfnamefont {R.~A.}\ \bibnamefont {Cox}},
  \bibinfo {author} {\bibfnamefont {J.~N.}\ \bibnamefont {Crowley}}, \bibinfo
  {author} {\bibfnamefont {R.~F.}\ \bibnamefont {Hampson}}, \bibinfo {author}
  {\bibfnamefont {R.~G.}\ \bibnamefont {Hynes}}, \bibinfo {author}
  {\bibfnamefont {M.~E.}\ \bibnamefont {Jenkin}}, \bibinfo {author}
  {\bibfnamefont {M.~J.}\ \bibnamefont {Rossi}}, \bibinfo {author}
  {\bibfnamefont {J.}~\bibnamefont {Troe}},\ and\ \bibinfo {author}
  {\bibfnamefont {I.}~\bibnamefont {Subcommittee}},\ }\bibfield  {title}
  {\enquote {\bibinfo {title} {Evaluated kinetic and photochemical data for
  atmospheric chemistry: Volume {II} - gas phase reactions of organic
  species},}\ }\href {https://doi.org/10.5194/acp-6-3625-2006} {\bibfield
  {journal} {\bibinfo  {journal} {Atmos. Chem. Phys.}\ }\textbf {\bibinfo
  {volume} {6}},\ \bibinfo {pages} {3625--4055} (\bibinfo {year}
  {2006})}\BibitemShut {NoStop}%
\bibitem [{\citenamefont {Atkinson}(1986)}]{Atkinson:CR86:69}%
  \BibitemOpen
  \bibfield  {author} {\bibinfo {author} {\bibfnamefont {R.}~\bibnamefont
  {Atkinson}},\ }\bibfield  {title} {\enquote {\bibinfo {title} {Kinetics and
  mechanisms of the gas-phase reactions of the hydroxyl radical with organic
  compounds under atmospheric conditions},}\ }\href
  {https://doi.org/10.1021/cr00071a004} {\bibfield  {journal} {\bibinfo
  {journal} {Chem. Rev.}\ }\textbf {\bibinfo {volume} {86}},\ \bibinfo {pages}
  {69--201} (\bibinfo {year} {1986})}\BibitemShut {NoStop}%
\bibitem [{\citenamefont {Aschmann}\ and\ \citenamefont
  {Atkinson}(1994)}]{Aschmann:EST28:1539}%
  \BibitemOpen
  \bibfield  {author} {\bibinfo {author} {\bibfnamefont {S.~M.}\ \bibnamefont
  {Aschmann}}\ and\ \bibinfo {author} {\bibfnamefont {R.}~\bibnamefont
  {Atkinson}},\ }\bibfield  {title} {\enquote {\bibinfo {title} {Formation
  yields of methyl vinyl ketone and methacrolein from the gas-phase reaction of
  {O}$_{3}$ with isoprene},}\ }\href {https://doi.org/10.1021/es00057a025}
  {\bibfield  {journal} {\bibinfo  {journal} {Environ. Sci. Technol.}\ }\textbf
  {\bibinfo {volume} {28}},\ \bibinfo {pages} {1539--1542} (\bibinfo {year}
  {1994})}\BibitemShut {NoStop}%
\bibitem [{\citenamefont {Kroll}\ and\ \citenamefont
  {Seinfeld}(2008)}]{Kroll:AE42:3593}%
  \BibitemOpen
  \bibfield  {author} {\bibinfo {author} {\bibfnamefont {J.~H.}\ \bibnamefont
  {Kroll}}\ and\ \bibinfo {author} {\bibfnamefont {J.~H.}\ \bibnamefont
  {Seinfeld}},\ }\bibfield  {title} {\enquote {\bibinfo {title} {Chemistry of
  secondary organic aerosol: Formation and evolution of low-volatility organics
  in the atmosphere},}\ }\href
  {https://doi.org/https://doi.org/10.1016/j.atmosenv.2008.01.003} {\bibfield
  {journal} {\bibinfo  {journal} {Atmos. Environ.}\ }\textbf {\bibinfo {volume}
  {42}},\ \bibinfo {pages} {3593 -- 3624} (\bibinfo {year} {2008})}\BibitemShut
  {NoStop}%
\bibitem [{\citenamefont {Jorgensen}, \citenamefont {Lim},\ and\ \citenamefont
  {Blake}(1993)}]{Jorgensen:JACS115:2936}%
  \BibitemOpen
  \bibfield  {author} {\bibinfo {author} {\bibfnamefont {W.~L.}\ \bibnamefont
  {Jorgensen}}, \bibinfo {author} {\bibfnamefont {D.}~\bibnamefont {Lim}},\
  and\ \bibinfo {author} {\bibfnamefont {J.~F.}\ \bibnamefont {Blake}},\
  }\bibfield  {title} {\enquote {\bibinfo {title} {Ab initio study of
  diels-alder reactions of cyclopentadiene with ethylene, isoprene,
  cyclopentadiene, acrylonitrile, and methyl vinyl ketone},}\ }\href
  {https://doi.org/10.1021/ja00060a048} {\bibfield  {journal} {\bibinfo
  {journal} {J. Am. Chem. Soc.}\ }\textbf {\bibinfo {volume} {115}},\ \bibinfo
  {pages} {2936--2942} (\bibinfo {year} {1993})}\BibitemShut {NoStop}%
\bibitem [{\citenamefont {Jorgensen}\ \emph {et~al.}(1994)\citenamefont
  {Jorgensen}, \citenamefont {Blake}, \citenamefont {Lim},\ and\ \citenamefont
  {Severance}}]{Jorgensen:JCSFT90:1727}%
  \BibitemOpen
  \bibfield  {author} {\bibinfo {author} {\bibfnamefont {W.~L.}\ \bibnamefont
  {Jorgensen}}, \bibinfo {author} {\bibfnamefont {J.~F.}\ \bibnamefont
  {Blake}}, \bibinfo {author} {\bibfnamefont {D.}~\bibnamefont {Lim}},\ and\
  \bibinfo {author} {\bibfnamefont {D.~L.}\ \bibnamefont {Severance}},\
  }\bibfield  {title} {\enquote {\bibinfo {title} {Investigation of solvent
  effects on pericyclic reactions by computer simulations},}\ }\href
  {https://doi.org/10.1039/FT9949001727} {\bibfield  {journal} {\bibinfo
  {journal} {J. Chem. Soc.{,} Faraday Trans.}\ }\textbf {\bibinfo {volume}
  {90}},\ \bibinfo {pages} {1727--1732} (\bibinfo {year} {1994})}\BibitemShut
  {NoStop}%
\bibitem [{\citenamefont {Rogers}(1947)}]{Rogers:JACS69:2544}%
  \BibitemOpen
  \bibfield  {author} {\bibinfo {author} {\bibfnamefont {M.~T.}\ \bibnamefont
  {Rogers}},\ }\bibfield  {title} {\enquote {\bibinfo {title} {The electric
  moments and ultraviolet absorption spectra of some derivatives of
  cyclopropane and of ethylene oxide},}\ }\href
  {https://doi.org/10.1021/ja01202a081} {\bibfield  {journal} {\bibinfo
  {journal} {J. Am. Chem. Soc.}\ }\textbf {\bibinfo {volume} {69}},\ \bibinfo
  {pages} {2544--2548} (\bibinfo {year} {1947})}\BibitemShut {NoStop}%
\bibitem [{\citenamefont {Foster}, \citenamefont {Rao},\ and\ \citenamefont
  {Curl~Jr}(1965)}]{Foster:JCP43:1064}%
  \BibitemOpen
  \bibfield  {author} {\bibinfo {author} {\bibfnamefont {P.~D.}\ \bibnamefont
  {Foster}}, \bibinfo {author} {\bibfnamefont {V.~M.}\ \bibnamefont {Rao}},\
  and\ \bibinfo {author} {\bibfnamefont {R.~F.}\ \bibnamefont {Curl~Jr}},\
  }\bibfield  {title} {\enquote {\bibinfo {title} {Microwave spectrum of methyl
  vinyl ketone},}\ }\href {https://doi.org/10.1063/1.1696821} {\bibfield
  {journal} {\bibinfo  {journal} {J. Chem. Phys.}\ }\textbf {\bibinfo {volume}
  {43}},\ \bibinfo {pages} {1064--1066} (\bibinfo {year} {1965})}\BibitemShut
  {NoStop}%
\bibitem [{\citenamefont {Fantoni}, \citenamefont {Caminati},\ and\
  \citenamefont {Meyer}(1987)}]{Fantoni:CPL133:27}%
  \BibitemOpen
  \bibfield  {author} {\bibinfo {author} {\bibfnamefont {A.}~\bibnamefont
  {Fantoni}}, \bibinfo {author} {\bibfnamefont {W.}~\bibnamefont {Caminati}},\
  and\ \bibinfo {author} {\bibfnamefont {R.}~\bibnamefont {Meyer}},\ }\bibfield
   {title} {\enquote {\bibinfo {title} {Torsional interactions in methyl vinyl
  ketone},}\ }\href {https://doi.org/10.1016/0009-2614(87)80047-7} {\bibfield
  {journal} {\bibinfo  {journal} {Chem. Phys. Lett.}\ }\textbf {\bibinfo
  {volume} {133}},\ \bibinfo {pages} {27--33} (\bibinfo {year}
  {1987})}\BibitemShut {NoStop}%
\bibitem [{\citenamefont {De~Smedt}\ \emph {et~al.}(1989)\citenamefont
  {De~Smedt}, \citenamefont {Vanhouteghem}, \citenamefont {Van~Alsenoy},
  \citenamefont {Geise}, \citenamefont {Van~der Veken},\ and\ \citenamefont
  {Coppens}}]{DeSmedt:JMS195:227}%
  \BibitemOpen
  \bibfield  {author} {\bibinfo {author} {\bibfnamefont {J.}~\bibnamefont
  {De~Smedt}}, \bibinfo {author} {\bibfnamefont {F.}~\bibnamefont
  {Vanhouteghem}}, \bibinfo {author} {\bibfnamefont {C.}~\bibnamefont
  {Van~Alsenoy}}, \bibinfo {author} {\bibfnamefont {H.}~\bibnamefont {Geise}},
  \bibinfo {author} {\bibfnamefont {B.}~\bibnamefont {Van~der Veken}},\ and\
  \bibinfo {author} {\bibfnamefont {P.}~\bibnamefont {Coppens}},\ }\bibfield
  {title} {\enquote {\bibinfo {title} {Methyl vinyl ketone in the gas phase,
  investigated by electron diffraction, infrared band contour analysis and
  microwave spectroscopy, supplemented with ab-initio calculations of
  geometries and force fields},}\ }\href
  {https://doi.org/10.1016/0022-2860(89)80171-1} {\bibfield  {journal}
  {\bibinfo  {journal} {J. Mol. Struct.}\ }\textbf {\bibinfo {volume} {195}},\
  \bibinfo {pages} {227--251} (\bibinfo {year} {1989})}\BibitemShut {NoStop}%
\bibitem [{\citenamefont {Noack}\ and\ \citenamefont
  {Jones}(1961)}]{Noack:CJC39:2225}%
  \BibitemOpen
  \bibfield  {author} {\bibinfo {author} {\bibfnamefont {K.}~\bibnamefont
  {Noack}}\ and\ \bibinfo {author} {\bibfnamefont {R.~N.}\ \bibnamefont
  {Jones}},\ }\bibfield  {title} {\enquote {\bibinfo {title} {Conformational
  equilibria in open-chain $\alpha$, $\beta$-unsaturated ketones},}\ }\href
  {https://doi.org/10.1139/v61-294} {\bibfield  {journal} {\bibinfo  {journal}
  {Can. J. Chem.}\ }\textbf {\bibinfo {volume} {39}},\ \bibinfo {pages}
  {2225--2235} (\bibinfo {year} {1961})}\BibitemShut {NoStop}%
\bibitem [{\citenamefont {Bowles}, \citenamefont {George},\ and\ \citenamefont
  {Maddams}(1969)}]{Bowles:JCSB:810}%
  \BibitemOpen
  \bibfield  {author} {\bibinfo {author} {\bibfnamefont {A.}~\bibnamefont
  {Bowles}}, \bibinfo {author} {\bibfnamefont {W.}~\bibnamefont {George}},\
  and\ \bibinfo {author} {\bibfnamefont {W.}~\bibnamefont {Maddams}},\
  }\bibfield  {title} {\enquote {\bibinfo {title} {Conformations of some
  $\alpha$, $\beta$-unsaturated carbonyl compounds. {P}art {I}. infrared
  spectra of acraldehyde, crotonaldehyde, methyl vinyl ketone, and
  ethylideneacetone},}\ }\href {https://doi.org/10.1039/j29690000810}
  {\bibfield  {journal} {\bibinfo  {journal} {J. Chem. Soc. B}\ ,\ \bibinfo
  {pages} {810--818}} (\bibinfo {year} {1969})}\BibitemShut {NoStop}%
\bibitem [{\citenamefont {Durig}\ and\ \citenamefont
  {Little}(1981)}]{Durig:JCP75:3660}%
  \BibitemOpen
  \bibfield  {author} {\bibinfo {author} {\bibfnamefont {J.}~\bibnamefont
  {Durig}}\ and\ \bibinfo {author} {\bibfnamefont {T.}~\bibnamefont {Little}},\
  }\bibfield  {title} {\enquote {\bibinfo {title} {Conformational barriers to
  internal rotation and vibrational assignment of methyl vinyl ketone},}\
  }\href {https://doi.org/10.1063/1.442530} {\bibfield  {journal} {\bibinfo
  {journal} {J. Chem. Phys.}\ }\textbf {\bibinfo {volume} {75}},\ \bibinfo
  {pages} {3660--3668} (\bibinfo {year} {1981})}\BibitemShut {NoStop}%
\bibitem [{\citenamefont {Oelichmann}, \citenamefont {Bougeard},\ and\
  \citenamefont {Schrader}(1981)}]{Oelichmann:JMS77:179}%
  \BibitemOpen
  \bibfield  {author} {\bibinfo {author} {\bibfnamefont {H.-J.}\ \bibnamefont
  {Oelichmann}}, \bibinfo {author} {\bibfnamefont {D.}~\bibnamefont
  {Bougeard}},\ and\ \bibinfo {author} {\bibfnamefont {B.}~\bibnamefont
  {Schrader}},\ }\bibfield  {title} {\enquote {\bibinfo {title} {Coupled
  calculation of vibrational frequencies and intensities: Part {VI}. {IR} and
  raman spectra of crotonaldehyde, methacrolein and methyl-vinylketone},}\
  }\href {https://doi.org/https://doi.org/10.1016/0022-2860(81)80063-4}
  {\bibfield  {journal} {\bibinfo  {journal} {J. Mol. Struct.}\ }\textbf
  {\bibinfo {volume} {77}},\ \bibinfo {pages} {179 -- 194} (\bibinfo {year}
  {1981})}\BibitemShut {NoStop}%
\bibitem [{\citenamefont {Wilcox}\ \emph {et~al.}(2011)\citenamefont {Wilcox},
  \citenamefont {Shirar}, \citenamefont {Williams},\ and\ \citenamefont
  {Dian}}]{Wilcox:CPL508:10}%
  \BibitemOpen
  \bibfield  {author} {\bibinfo {author} {\bibfnamefont {D.~S.}\ \bibnamefont
  {Wilcox}}, \bibinfo {author} {\bibfnamefont {A.~J.}\ \bibnamefont {Shirar}},
  \bibinfo {author} {\bibfnamefont {O.~L.}\ \bibnamefont {Williams}},\ and\
  \bibinfo {author} {\bibfnamefont {B.~C.}\ \bibnamefont {Dian}},\ }\bibfield
  {title} {\enquote {\bibinfo {title} {Additional conformer observed in the
  microwave spectrum of methyl vinyl ketone},}\ }\href
  {https://doi.org/10.1016/j.cplett.2011.04.001} {\bibfield  {journal}
  {\bibinfo  {journal} {Chem. Phys. Lett.}\ }\textbf {\bibinfo {volume}
  {508}},\ \bibinfo {pages} {10--16} (\bibinfo {year} {2011})}\BibitemShut
  {NoStop}%
\bibitem [{\citenamefont {Zakharenko}\ \emph {et~al.}(2017)\citenamefont
  {Zakharenko}, \citenamefont {Motiyenko}, \citenamefont {Aviles~Moreno},\ and\
  \citenamefont {Huet}}]{Zakharenko:JPCA121:6420}%
  \BibitemOpen
  \bibfield  {author} {\bibinfo {author} {\bibfnamefont {O.}~\bibnamefont
  {Zakharenko}}, \bibinfo {author} {\bibfnamefont {R.~A.}\ \bibnamefont
  {Motiyenko}}, \bibinfo {author} {\bibfnamefont {J.~R.}\ \bibnamefont
  {Aviles~Moreno}},\ and\ \bibinfo {author} {\bibfnamefont {T.~R.}\
  \bibnamefont {Huet}},\ }\bibfield  {title} {\enquote {\bibinfo {title}
  {Conformational landscape and torsion–rotation–vibration effects in the
  two conformers of methyl vinyl ketone, a major oxidation product of
  isoprene},}\ }\href {https://doi.org/10.1021/acs.jpca.7b06360} {\bibfield
  {journal} {\bibinfo  {journal} {J. Phys. Chem. A}\ }\textbf {\bibinfo
  {volume} {121}},\ \bibinfo {pages} {6420--6428} (\bibinfo {year}
  {2017})}\BibitemShut {NoStop}%
\bibitem [{\citenamefont {Lindenmaier}\ \emph {et~al.}(2017)\citenamefont
  {Lindenmaier}, \citenamefont {Williams}, \citenamefont {Sams},\ and\
  \citenamefont {Johnson}}]{Lindenmaier:JPCA121:1195}%
  \BibitemOpen
  \bibfield  {author} {\bibinfo {author} {\bibfnamefont {R.}~\bibnamefont
  {Lindenmaier}}, \bibinfo {author} {\bibfnamefont {S.~D.}\ \bibnamefont
  {Williams}}, \bibinfo {author} {\bibfnamefont {R.~L.}\ \bibnamefont {Sams}},\
  and\ \bibinfo {author} {\bibfnamefont {T.~J.}\ \bibnamefont {Johnson}},\
  }\bibfield  {title} {\enquote {\bibinfo {title} {Quantitative infrared
  absorption spectra and vibrational assignments of crotonaldehyde and methyl
  vinyl ketone using gas-phase mid-infrared, far-infrared, and liquid raman
  spectra: \textit{s-cis} vs \textit{s-trans} composition confirmed via
  temperature studies and ab initio methods},}\ }\href
  {https://doi.org/10.1021/acs.jpca.6b10872} {\bibfield  {journal} {\bibinfo
  {journal} {J. Phys. Chem. A}\ }\textbf {\bibinfo {volume} {121}},\ \bibinfo
  {pages} {1195--1212} (\bibinfo {year} {2017})}\BibitemShut {NoStop}%
\bibitem [{\citenamefont {Chang}\ \emph {et~al.}(2013)\citenamefont {Chang},
  \citenamefont {D{\l}ugo\l\k{e}cki}, \citenamefont {K{\"u}pper}, \citenamefont
  {R{\"o}sch}, \citenamefont {Wild},\ and\ \citenamefont
  {Willitsch}}]{Chang:Science342:98}%
  \BibitemOpen
  \bibfield  {author} {\bibinfo {author} {\bibfnamefont {Y.-P.}\ \bibnamefont
  {Chang}}, \bibinfo {author} {\bibfnamefont {K.}~\bibnamefont
  {D{\l}ugo\l\k{e}cki}}, \bibinfo {author} {\bibfnamefont {J.}~\bibnamefont
  {K{\"u}pper}}, \bibinfo {author} {\bibfnamefont {D.}~\bibnamefont
  {R{\"o}sch}}, \bibinfo {author} {\bibfnamefont {D.}~\bibnamefont {Wild}},\
  and\ \bibinfo {author} {\bibfnamefont {S.}~\bibnamefont {Willitsch}},\
  }\bibfield  {title} {\enquote {\bibinfo {title} {Specific chemical
  reactivities of spatially separated 3-aminophenol conformers with cold
  {Ca$^+$} ions},}\ }\href {https://doi.org/10.1126/science.1242271} {\bibfield
   {journal} {\bibinfo  {journal} {Science}\ }\textbf {\bibinfo {volume}
  {342}},\ \bibinfo {pages} {98--101} (\bibinfo {year} {2013})},\ \Eprint
  {https://arxiv.org/abs/1308.6538} {arXiv:1308.6538 [physics]} \BibitemShut
  {NoStop}%
\bibitem [{\citenamefont {Willitsch}(2017)}]{Willitsch:ACP162:307}%
  \BibitemOpen
  \bibfield  {author} {\bibinfo {author} {\bibfnamefont {S.}~\bibnamefont
  {Willitsch}},\ }\bibfield  {title} {\enquote {\bibinfo {title} {Chemistry
  with controlled ions},}\ }\href {https://doi.org/10.1002/9781119324560.ch5}
  {\bibfield  {journal} {\bibinfo  {journal} {Adv. Chem. Phys.}\ }\textbf
  {\bibinfo {volume} {162}},\ \bibinfo {pages} {307} (\bibinfo {year}
  {2017})}\BibitemShut {NoStop}%
\bibitem [{\citenamefont {Chang}\ \emph {et~al.}(2015)\citenamefont {Chang},
  \citenamefont {Horke}, \citenamefont {Trippel},\ and\ \citenamefont
  {Küpper}}]{Chang:IRPC34:557}%
  \BibitemOpen
  \bibfield  {author} {\bibinfo {author} {\bibfnamefont {Y.-P.}\ \bibnamefont
  {Chang}}, \bibinfo {author} {\bibfnamefont {D.~A.}\ \bibnamefont {Horke}},
  \bibinfo {author} {\bibfnamefont {S.}~\bibnamefont {Trippel}},\ and\ \bibinfo
  {author} {\bibfnamefont {J.}~\bibnamefont {Küpper}},\ }\bibfield  {title}
  {\enquote {\bibinfo {title} {Spatially-controlled complex molecules and their
  applications},}\ }\href {https://doi.org/10.1080/0144235X.2015.1077838}
  {\bibfield  {journal} {\bibinfo  {journal} {Int. Rev. Phys. Chem.}\ }\textbf
  {\bibinfo {volume} {34}},\ \bibinfo {pages} {557--590} (\bibinfo {year}
  {2015})},\ \Eprint {https://arxiv.org/abs/1505.05632} {arXiv:1505.05632
  [physics]} \BibitemShut {NoStop}%
\bibitem [{\citenamefont {Nielsen}\ \emph {et~al.}(2011)\citenamefont
  {Nielsen}, \citenamefont {Simesen}, \citenamefont {Bisgaard}, \citenamefont
  {Stapelfeldt}, \citenamefont {Filsinger}, \citenamefont {Friedrich},
  \citenamefont {Meijer},\ and\ \citenamefont
  {K{\"u}pper}}]{Nielsen:PCCP13:18971}%
  \BibitemOpen
  \bibfield  {author} {\bibinfo {author} {\bibfnamefont {J.~H.}\ \bibnamefont
  {Nielsen}}, \bibinfo {author} {\bibfnamefont {P.}~\bibnamefont {Simesen}},
  \bibinfo {author} {\bibfnamefont {C.~Z.}\ \bibnamefont {Bisgaard}}, \bibinfo
  {author} {\bibfnamefont {H.}~\bibnamefont {Stapelfeldt}}, \bibinfo {author}
  {\bibfnamefont {F.}~\bibnamefont {Filsinger}}, \bibinfo {author}
  {\bibfnamefont {B.}~\bibnamefont {Friedrich}}, \bibinfo {author}
  {\bibfnamefont {G.}~\bibnamefont {Meijer}},\ and\ \bibinfo {author}
  {\bibfnamefont {J.}~\bibnamefont {K{\"u}pper}},\ }\bibfield  {title}
  {\enquote {\bibinfo {title} {Stark-selected beam of ground-state {OCS}
  molecules characterized by revivals of impulsive alignment},}\ }\href
  {https://doi.org/10.1039/c1cp21143a} {\bibfield  {journal} {\bibinfo
  {journal} {Phys. Chem. Chem. Phys.}\ }\textbf {\bibinfo {volume} {13}},\
  \bibinfo {pages} {18971--18975} (\bibinfo {year} {2011})},\ \Eprint
  {https://arxiv.org/abs/1105.2413} {arXiv:1105.2413 [physics]} \BibitemShut
  {NoStop}%
\bibitem [{\citenamefont {Horke}\ \emph {et~al.}(2014)\citenamefont {Horke},
  \citenamefont {Chang}, \citenamefont {D{\l}ugo{\l}\k{e}cki},\ and\
  \citenamefont {K{\"u}pper}}]{Horke:ACIE53:11965}%
  \BibitemOpen
  \bibfield  {author} {\bibinfo {author} {\bibfnamefont {D.~A.}\ \bibnamefont
  {Horke}}, \bibinfo {author} {\bibfnamefont {Y.-P.}\ \bibnamefont {Chang}},
  \bibinfo {author} {\bibfnamefont {K.}~\bibnamefont {D{\l}ugo{\l}\k{e}cki}},\
  and\ \bibinfo {author} {\bibfnamefont {J.}~\bibnamefont {K{\"u}pper}},\
  }\bibfield  {title} {\enquote {\bibinfo {title} {Separating para and ortho
  water},}\ }\href {https://doi.org/10.1002/anie.201405986} {\bibfield
  {journal} {\bibinfo  {journal} {Angew. Chem. Int. Ed.}\ }\textbf {\bibinfo
  {volume} {53}},\ \bibinfo {pages} {11965--11968} (\bibinfo {year} {2014})},\
  \Eprint {https://arxiv.org/abs/1407.2056} {arXiv:1407.2056 [physics]}
  \BibitemShut {NoStop}%
\bibitem [{\citenamefont {Kienitz}\ \emph {et~al.}(2017)\citenamefont
  {Kienitz}, \citenamefont {D{\l}ugo{\l}\k{e}cki}, \citenamefont {Trippel},\
  and\ \citenamefont {Küpper}}]{Kienitz:JCP147:024304}%
  \BibitemOpen
  \bibfield  {author} {\bibinfo {author} {\bibfnamefont {J.~S.}\ \bibnamefont
  {Kienitz}}, \bibinfo {author} {\bibfnamefont {K.}~\bibnamefont
  {D{\l}ugo{\l}\k{e}cki}}, \bibinfo {author} {\bibfnamefont {S.}~\bibnamefont
  {Trippel}},\ and\ \bibinfo {author} {\bibfnamefont {J.}~\bibnamefont
  {Küpper}},\ }\bibfield  {title} {\enquote {\bibinfo {title} {Improved
  spatial separation of neutral molecules},}\ }\href
  {https://doi.org/10.1063/1.4991479} {\bibfield  {journal} {\bibinfo
  {journal} {J. Chem. Phys.}\ }\textbf {\bibinfo {volume} {147}},\ \bibinfo
  {pages} {024304} (\bibinfo {year} {2017})},\ \Eprint
  {https://arxiv.org/abs/1704.08912} {arXiv:1704.08912 [physics]} \BibitemShut
  {NoStop}%
\bibitem [{\citenamefont {Filsinger}\ \emph {et~al.}(2008)\citenamefont
  {Filsinger}, \citenamefont {Erlekam}, \citenamefont {von Helden},
  \citenamefont {K{\"u}pper},\ and\ \citenamefont
  {Meijer}}]{Filsinger:PRL100:133003}%
  \BibitemOpen
  \bibfield  {author} {\bibinfo {author} {\bibfnamefont {F.}~\bibnamefont
  {Filsinger}}, \bibinfo {author} {\bibfnamefont {U.}~\bibnamefont {Erlekam}},
  \bibinfo {author} {\bibfnamefont {G.}~\bibnamefont {von Helden}}, \bibinfo
  {author} {\bibfnamefont {J.}~\bibnamefont {K{\"u}pper}},\ and\ \bibinfo
  {author} {\bibfnamefont {G.}~\bibnamefont {Meijer}},\ }\bibfield  {title}
  {\enquote {\bibinfo {title} {Selector for structural isomers of neutral
  molecules},}\ }\href {https://doi.org/10.1103/PhysRevLett.100.133003}
  {\bibfield  {journal} {\bibinfo  {journal} {Phys. Rev. Lett.}\ }\textbf
  {\bibinfo {volume} {100}},\ \bibinfo {pages} {133003} (\bibinfo {year}
  {2008})},\ \Eprint {https://arxiv.org/abs/0802.2795} {arXiv:0802.2795
  [physics]} \BibitemShut {NoStop}%
\bibitem [{\citenamefont {Filsinger}\ \emph
  {et~al.}(2009{\natexlab{a}})\citenamefont {Filsinger}, \citenamefont
  {K{\"u}pper}, \citenamefont {Meijer}, \citenamefont {Hansen}, \citenamefont
  {Maurer}, \citenamefont {Nielsen}, \citenamefont {Holmegaard},\ and\
  \citenamefont {Stapelfeldt}}]{Filsinger:ACIE48:6900}%
  \BibitemOpen
  \bibfield  {author} {\bibinfo {author} {\bibfnamefont {F.}~\bibnamefont
  {Filsinger}}, \bibinfo {author} {\bibfnamefont {J.}~\bibnamefont
  {K{\"u}pper}}, \bibinfo {author} {\bibfnamefont {G.}~\bibnamefont {Meijer}},
  \bibinfo {author} {\bibfnamefont {J.~L.}\ \bibnamefont {Hansen}}, \bibinfo
  {author} {\bibfnamefont {J.}~\bibnamefont {Maurer}}, \bibinfo {author}
  {\bibfnamefont {J.~H.}\ \bibnamefont {Nielsen}}, \bibinfo {author}
  {\bibfnamefont {L.}~\bibnamefont {Holmegaard}},\ and\ \bibinfo {author}
  {\bibfnamefont {H.}~\bibnamefont {Stapelfeldt}},\ }\bibfield  {title}
  {\enquote {\bibinfo {title} {Pure samples of individual conformers: the
  separation of stereo-isomers of complex molecules using electric fields},}\
  }\href {https://doi.org/10.1002/anie.200902650} {\bibfield  {journal}
  {\bibinfo  {journal} {Angew. Chem. Int. Ed.}\ }\textbf {\bibinfo {volume}
  {48}},\ \bibinfo {pages} {6900--6902} (\bibinfo {year}
  {2009}{\natexlab{a}})}\BibitemShut {NoStop}%
\bibitem [{\citenamefont {Kierspel}\ \emph {et~al.}(2014)\citenamefont
  {Kierspel}, \citenamefont {Horke}, \citenamefont {Chang},\ and\ \citenamefont
  {K{\"u}pper}}]{Kierspel:CPL591:130}%
  \BibitemOpen
  \bibfield  {author} {\bibinfo {author} {\bibfnamefont {T.}~\bibnamefont
  {Kierspel}}, \bibinfo {author} {\bibfnamefont {D.~A.}\ \bibnamefont {Horke}},
  \bibinfo {author} {\bibfnamefont {Y.-P.}\ \bibnamefont {Chang}},\ and\
  \bibinfo {author} {\bibfnamefont {J.}~\bibnamefont {K{\"u}pper}},\ }\bibfield
   {title} {\enquote {\bibinfo {title} {Spatially separated polar samples of
  the \emph{cis} and \emph{trans} conformers of 3-fluorophenol},}\ }\href
  {https://doi.org/10.1016/j.cplett.2013.11.010} {\bibfield  {journal}
  {\bibinfo  {journal} {Chem. Phys. Lett.}\ }\textbf {\bibinfo {volume}
  {591}},\ \bibinfo {pages} {130--132} (\bibinfo {year} {2014})},\ \Eprint
  {https://arxiv.org/abs/1312.4417} {arXiv:1312.4417 [physics]} \BibitemShut
  {NoStop}%
\bibitem [{\citenamefont {Teschmit}, \citenamefont {Horke},\ and\ \citenamefont
  {K\"{u}pper}(2018)}]{Teschmit:ACIE57:13775}%
  \BibitemOpen
  \bibfield  {author} {\bibinfo {author} {\bibfnamefont {N.}~\bibnamefont
  {Teschmit}}, \bibinfo {author} {\bibfnamefont {D.~A.}\ \bibnamefont
  {Horke}},\ and\ \bibinfo {author} {\bibfnamefont {J.}~\bibnamefont
  {K\"{u}pper}},\ }\bibfield  {title} {\enquote {\bibinfo {title} {Spatially
  separating the conformers of a dipeptide},}\ }\href
  {https://doi.org/10.1002/anie.201807646} {\bibfield  {journal} {\bibinfo
  {journal} {Angew. Chem. Int. Ed.}\ }\textbf {\bibinfo {volume} {57}},\
  \bibinfo {pages} {13775--13779} (\bibinfo {year} {2018})},\ \Eprint
  {https://arxiv.org/abs/1805.12396} {arXiv:1805.12396 [physics]} \BibitemShut
  {NoStop}%
\bibitem [{\citenamefont {You}\ \emph {et~al.}(2018)\citenamefont {You},
  \citenamefont {Kim}, \citenamefont {Han}, \citenamefont {Ahn}, \citenamefont
  {Lim},\ and\ \citenamefont {Kim}}]{You:JPCA122:1194}%
  \BibitemOpen
  \bibfield  {author} {\bibinfo {author} {\bibfnamefont {H.~S.}\ \bibnamefont
  {You}}, \bibinfo {author} {\bibfnamefont {J.}~\bibnamefont {Kim}}, \bibinfo
  {author} {\bibfnamefont {S.}~\bibnamefont {Han}}, \bibinfo {author}
  {\bibfnamefont {D.-S.}\ \bibnamefont {Ahn}}, \bibinfo {author} {\bibfnamefont
  {J.~S.}\ \bibnamefont {Lim}},\ and\ \bibinfo {author} {\bibfnamefont {S.~K.}\
  \bibnamefont {Kim}},\ }\bibfield  {title} {\enquote {\bibinfo {title}
  {Spatial isolation of conformational isomers of hydroquinone and its water
  cluster using the stark deflector},}\ }\href
  {https://doi.org/10.1021/acs.jpca.7b10431} {\bibfield  {journal} {\bibinfo
  {journal} {J. Phys. Chem. A}\ }\textbf {\bibinfo {volume} {122}},\ \bibinfo
  {pages} {1194} (\bibinfo {year} {2018})}\BibitemShut {NoStop}%
\bibitem [{\citenamefont {Trippel}\ \emph {et~al.}(2012)\citenamefont
  {Trippel}, \citenamefont {Chang}, \citenamefont {Stern}, \citenamefont
  {Mullins}, \citenamefont {Holmegaard},\ and\ \citenamefont
  {K{\"u}pper}}]{Trippel:PRA86:033202}%
  \BibitemOpen
  \bibfield  {author} {\bibinfo {author} {\bibfnamefont {S.}~\bibnamefont
  {Trippel}}, \bibinfo {author} {\bibfnamefont {Y.-P.}\ \bibnamefont {Chang}},
  \bibinfo {author} {\bibfnamefont {S.}~\bibnamefont {Stern}}, \bibinfo
  {author} {\bibfnamefont {T.}~\bibnamefont {Mullins}}, \bibinfo {author}
  {\bibfnamefont {L.}~\bibnamefont {Holmegaard}},\ and\ \bibinfo {author}
  {\bibfnamefont {J.}~\bibnamefont {K{\"u}pper}},\ }\bibfield  {title}
  {\enquote {\bibinfo {title} {Spatial separation of state- and size-selected
  neutral clusters},}\ }\href {https://doi.org/10.1103/PhysRevA.86.033202}
  {\bibfield  {journal} {\bibinfo  {journal} {Phys. Rev. A}\ }\textbf {\bibinfo
  {volume} {86}},\ \bibinfo {pages} {033202} (\bibinfo {year} {2012})},\
  \Eprint {https://arxiv.org/abs/1208.4935} {arXiv:1208.4935 [physics]}
  \BibitemShut {NoStop}%
\bibitem [{\citenamefont {Johny}\ \emph {et~al.}(2019)\citenamefont {Johny},
  \citenamefont {Onvlee}, \citenamefont {Kierspel}, \citenamefont {Bieker},
  \citenamefont {Trippel},\ and\ \citenamefont {Küpper}}]{Johny:CPL721:149}%
  \BibitemOpen
  \bibfield  {author} {\bibinfo {author} {\bibfnamefont {M.}~\bibnamefont
  {Johny}}, \bibinfo {author} {\bibfnamefont {J.}~\bibnamefont {Onvlee}},
  \bibinfo {author} {\bibfnamefont {T.}~\bibnamefont {Kierspel}}, \bibinfo
  {author} {\bibfnamefont {H.}~\bibnamefont {Bieker}}, \bibinfo {author}
  {\bibfnamefont {S.}~\bibnamefont {Trippel}},\ and\ \bibinfo {author}
  {\bibfnamefont {J.}~\bibnamefont {Küpper}},\ }\bibfield  {title} {\enquote
  {\bibinfo {title} {Spatial separation of pyrrole and pyrrole-water
  clusters},}\ }\href {https://doi.org/10.1016/j.cplett.2019.01.052} {\bibfield
   {journal} {\bibinfo  {journal} {Chem. Phys. Lett.}\ }\textbf {\bibinfo
  {volume} {721}},\ \bibinfo {pages} {149–152} (\bibinfo {year} {2019})},\
  \Eprint {https://arxiv.org/abs/1901.05267} {arXiv:1901.05267 [physics]}
  \BibitemShut {NoStop}%
\bibitem [{\citenamefont {Kilaj}\ \emph {et~al.}(2018)\citenamefont {Kilaj},
  \citenamefont {Gao}, \citenamefont {R\"osch}, \citenamefont {Rivero},
  \citenamefont {K\"upper},\ and\ \citenamefont
  {Willitsch}}]{Kilaj:NatComm9:2096}%
  \BibitemOpen
  \bibfield  {author} {\bibinfo {author} {\bibfnamefont {A.}~\bibnamefont
  {Kilaj}}, \bibinfo {author} {\bibfnamefont {H.}~\bibnamefont {Gao}}, \bibinfo
  {author} {\bibfnamefont {D.}~\bibnamefont {R\"osch}}, \bibinfo {author}
  {\bibfnamefont {U.}~\bibnamefont {Rivero}}, \bibinfo {author} {\bibfnamefont
  {J.}~\bibnamefont {K\"upper}},\ and\ \bibinfo {author} {\bibfnamefont
  {S.}~\bibnamefont {Willitsch}},\ }\bibfield  {title} {\enquote {\bibinfo
  {title} {Observation of different reactivities of para- and ortho-water
  towards trapped diazenylium ions},}\ }\href
  {https://doi.org/10.1038/s41467-018-04483-3} {\bibfield  {journal} {\bibinfo
  {journal} {Nat. Commun.}\ }\textbf {\bibinfo {volume} {9}},\ \bibinfo {pages}
  {2096} (\bibinfo {year} {2018})}\BibitemShut {NoStop}%
\bibitem [{\citenamefont {R{\"o}sch}\ \emph {et~al.}(2014)\citenamefont
  {R{\"o}sch}, \citenamefont {Willitsch}, \citenamefont {Chang},\ and\
  \citenamefont {K{\"u}pper}}]{Roesch:JCP140:124202}%
  \BibitemOpen
  \bibfield  {author} {\bibinfo {author} {\bibfnamefont {D.}~\bibnamefont
  {R{\"o}sch}}, \bibinfo {author} {\bibfnamefont {S.}~\bibnamefont
  {Willitsch}}, \bibinfo {author} {\bibfnamefont {Y.-P.}\ \bibnamefont
  {Chang}},\ and\ \bibinfo {author} {\bibfnamefont {J.}~\bibnamefont
  {K{\"u}pper}},\ }\bibfield  {title} {\enquote {\bibinfo {title} {Chemical
  reactions of conformationally selected 3-aminophenol molecules in a beam with
  coulomb-crystallized ca$^+$ ions},}\ }\href
  {https://doi.org/10.1063/1.4869100} {\bibfield  {journal} {\bibinfo
  {journal} {J. Chem. Phys.}\ }\textbf {\bibinfo {volume} {140}},\ \bibinfo
  {pages} {124202} (\bibinfo {year} {2014})}\BibitemShut {NoStop}%
\bibitem [{\citenamefont {Kilaj}\ \emph {et~al.}(2020)\citenamefont {Kilaj},
  \citenamefont {Gao}, \citenamefont {Tahchieva}, \citenamefont {Ramakrishnan},
  \citenamefont {Bachmann}, \citenamefont {Gillingham}, \citenamefont {von
  Lilienfeld}, \citenamefont {Küpper},\ and\ \citenamefont
  {Willitsch}}]{Kilaj:PCCP:inprep}%
  \BibitemOpen
  \bibfield  {author} {\bibinfo {author} {\bibfnamefont {A.}~\bibnamefont
  {Kilaj}}, \bibinfo {author} {\bibfnamefont {H.}~\bibnamefont {Gao}}, \bibinfo
  {author} {\bibfnamefont {D.}~\bibnamefont {Tahchieva}}, \bibinfo {author}
  {\bibfnamefont {R.}~\bibnamefont {Ramakrishnan}}, \bibinfo {author}
  {\bibfnamefont {D.}~\bibnamefont {Bachmann}}, \bibinfo {author}
  {\bibfnamefont {D.}~\bibnamefont {Gillingham}}, \bibinfo {author}
  {\bibfnamefont {O.~A.}\ \bibnamefont {von Lilienfeld}}, \bibinfo {author}
  {\bibfnamefont {J.}~\bibnamefont {Küpper}},\ and\ \bibinfo {author}
  {\bibfnamefont {S.}~\bibnamefont {Willitsch}},\ }\bibfield  {title} {\enquote
  {\bibinfo {title} {Quantum-chemistry-aided identification, synthesis and
  experimental validation of model systems for conformationally controlled
  reaction studies: Separation of the conformers of 2,3-dibromobuta-1,3-diene
  in the gas phase},}\ }\href {https://doi.org/10.1039/d0cp01396j} {\bibfield
  {journal} {\bibinfo  {journal} {Phys. Chem. Chem. Phys.}\ }\textbf {\bibinfo
  {volume} {22}},\ \bibinfo {pages} {13431--13439} (\bibinfo {year} {2020})},\
  \Eprint {https://arxiv.org/abs/2004.09659} {arXiv:2004.09659 [physics]}
  \BibitemShut {NoStop}%
\bibitem [{\citenamefont {K{\"u}pper}\ \emph {et~al.}(2014)\citenamefont
  {K{\"u}pper}, \citenamefont {Stern}, \citenamefont {Holmegaard},
  \citenamefont {Filsinger}, \citenamefont {Rouz\'{e}e}, \citenamefont
  {Rudenko}, \citenamefont {Johnsson}, \citenamefont {Martin}, \citenamefont
  {Adolph}, \citenamefont {Aquila} \emph {et~al.}}]{Kuepper:PRL112:083002}%
  \BibitemOpen
  \bibfield  {author} {\bibinfo {author} {\bibfnamefont {J.}~\bibnamefont
  {K{\"u}pper}}, \bibinfo {author} {\bibfnamefont {S.}~\bibnamefont {Stern}},
  \bibinfo {author} {\bibfnamefont {L.}~\bibnamefont {Holmegaard}}, \bibinfo
  {author} {\bibfnamefont {F.}~\bibnamefont {Filsinger}}, \bibinfo {author}
  {\bibfnamefont {A.}~\bibnamefont {Rouz\'{e}e}}, \bibinfo {author}
  {\bibfnamefont {A.}~\bibnamefont {Rudenko}}, \bibinfo {author} {\bibfnamefont
  {P.}~\bibnamefont {Johnsson}}, \bibinfo {author} {\bibfnamefont {A.~V.}\
  \bibnamefont {Martin}}, \bibinfo {author} {\bibfnamefont {M.}~\bibnamefont
  {Adolph}}, \bibinfo {author} {\bibfnamefont {A.}~\bibnamefont {Aquila}},
  \emph {et~al.},\ }\bibfield  {title} {\enquote {\bibinfo {title} {X-ray
  diffraction from isolated and strongly aligned gas-phase molecules with a
  free-electron laser},}\ }\href
  {https://doi.org/10.1103/PhysRevLett.112.083002} {\bibfield  {journal}
  {\bibinfo  {journal} {Phys. Rev. Lett.}\ }\textbf {\bibinfo {volume} {112}},\
  \bibinfo {pages} {083002} (\bibinfo {year} {2014})},\ \Eprint
  {https://arxiv.org/abs/1307.4577} {arXiv:1307.4577 [physics]} \BibitemShut
  {NoStop}%
\bibitem [{\citenamefont {Hensley}, \citenamefont {Yang},\ and\ \citenamefont
  {Centurion}(2012)}]{Hensley:PRL109:133202}%
  \BibitemOpen
  \bibfield  {author} {\bibinfo {author} {\bibfnamefont {C.~J.}\ \bibnamefont
  {Hensley}}, \bibinfo {author} {\bibfnamefont {J.}~\bibnamefont {Yang}},\ and\
  \bibinfo {author} {\bibfnamefont {M.}~\bibnamefont {Centurion}},\ }\bibfield
  {title} {\enquote {\bibinfo {title} {Imaging of isolated molecules with
  ultrafast electron pulses},}\ }\href
  {https://doi.org/10.1103/PhysRevLett.109.133202} {\bibfield  {journal}
  {\bibinfo  {journal} {Phys. Rev. Lett.}\ }\textbf {\bibinfo {volume} {109}},\
  \bibinfo {pages} {133202} (\bibinfo {year} {2012})}\BibitemShut {NoStop}%
\bibitem [{\citenamefont {Wiese}\ \emph {et~al.}(2020)\citenamefont {Wiese},
  \citenamefont {Onvlee}, \citenamefont {Trippel},\ and\ \citenamefont
  {Küpper}}]{Wiese:PRR2020:inprep}%
  \BibitemOpen
  \bibfield  {author} {\bibinfo {author} {\bibfnamefont {J.}~\bibnamefont
  {Wiese}}, \bibinfo {author} {\bibfnamefont {J.}~\bibnamefont {Onvlee}},
  \bibinfo {author} {\bibfnamefont {S.}~\bibnamefont {Trippel}},\ and\ \bibinfo
  {author} {\bibfnamefont {J.}~\bibnamefont {Küpper}},\ }\href@noop {}
  {\enquote {\bibinfo {title} {Strong-field ionization of complex molecules},}\
  } (\bibinfo {year} {2020}),\ \bibinfo {note} {under review},\ \Eprint
  {https://arxiv.org/abs/2003.02116} {arXiv:2003.02116 [physics]} \BibitemShut
  {NoStop}%
\bibitem [{\citenamefont {Wang}\ \emph {et~al.}(2020)\citenamefont {Wang},
  \citenamefont {He}, \citenamefont {Petrovic}, \citenamefont {Al-Refaie},
  \citenamefont {Bieker}, \citenamefont {Onvlee}, \citenamefont
  {Długołęcki},\ and\ \citenamefont {Küpper}}]{Wang:JMS1208:127863}%
  \BibitemOpen
  \bibfield  {author} {\bibinfo {author} {\bibfnamefont {J.}~\bibnamefont
  {Wang}}, \bibinfo {author} {\bibfnamefont {L.}~\bibnamefont {He}}, \bibinfo
  {author} {\bibfnamefont {J.}~\bibnamefont {Petrovic}}, \bibinfo {author}
  {\bibfnamefont {A.}~\bibnamefont {Al-Refaie}}, \bibinfo {author}
  {\bibfnamefont {H.}~\bibnamefont {Bieker}}, \bibinfo {author} {\bibfnamefont
  {J.}~\bibnamefont {Onvlee}}, \bibinfo {author} {\bibfnamefont
  {K.}~\bibnamefont {Długołęcki}},\ and\ \bibinfo {author} {\bibfnamefont
  {J.}~\bibnamefont {Küpper}},\ }\bibfield  {title} {\enquote {\bibinfo
  {title} {Spatial separation of 2-propanol monomer and its
  ionization-fragmentation pathways},}\ }\href
  {https://doi.org/https://doi.org/10.1016/j.molstruc.2020.127863} {\bibfield
  {journal} {\bibinfo  {journal} {J. Mol. Struct.}\ }\textbf {\bibinfo {volume}
  {1208}},\ \bibinfo {pages} {127863} (\bibinfo {year} {2020})}\BibitemShut
  {NoStop}%
\bibitem [{\citenamefont {Irimia}\ \emph {et~al.}(2009)\citenamefont {Irimia},
  \citenamefont {Dobrikov}, \citenamefont {Kortekaas}, \citenamefont {Voet},
  \citenamefont {van~den Ende}, \citenamefont {Groen},\ and\ \citenamefont
  {Janssen}}]{Irimia:RSI80:113303}%
  \BibitemOpen
  \bibfield  {author} {\bibinfo {author} {\bibfnamefont {D.}~\bibnamefont
  {Irimia}}, \bibinfo {author} {\bibfnamefont {D.}~\bibnamefont {Dobrikov}},
  \bibinfo {author} {\bibfnamefont {R.}~\bibnamefont {Kortekaas}}, \bibinfo
  {author} {\bibfnamefont {H.}~\bibnamefont {Voet}}, \bibinfo {author}
  {\bibfnamefont {D.~A.}\ \bibnamefont {van~den Ende}}, \bibinfo {author}
  {\bibfnamefont {W.~A.}\ \bibnamefont {Groen}},\ and\ \bibinfo {author}
  {\bibfnamefont {M.~H.~M.}\ \bibnamefont {Janssen}},\ }\bibfield  {title}
  {\enquote {\bibinfo {title} {A short pulse (7~$\mu$s {FWHM}) and high
  repetition rate (dc--5{kHz}) cantilever piezovalve for pulsed atomic and
  molecular beams},}\ }\href {https://doi.org/10.1063/1.3263912} {\bibfield
  {journal} {\bibinfo  {journal} {Rev. Sci. Instrum.}\ }\textbf {\bibinfo
  {volume} {80}},\ \bibinfo {pages} {113303} (\bibinfo {year}
  {2009})}\BibitemShut {NoStop}%
\bibitem [{\citenamefont {Kienitz}\ \emph {et~al.}(2016)\citenamefont
  {Kienitz}, \citenamefont {Trippel}, \citenamefont {Mullins}, \citenamefont
  {D{\l}ugo{\l}\k{e}cki}, \citenamefont {Gonz{\'a}lez-F{\'e}rez},\ and\
  \citenamefont {K\"upper}}]{Kienitz:CPC17:3740}%
  \BibitemOpen
  \bibfield  {author} {\bibinfo {author} {\bibfnamefont {J.~S.}\ \bibnamefont
  {Kienitz}}, \bibinfo {author} {\bibfnamefont {S.}~\bibnamefont {Trippel}},
  \bibinfo {author} {\bibfnamefont {T.}~\bibnamefont {Mullins}}, \bibinfo
  {author} {\bibfnamefont {K.}~\bibnamefont {D{\l}ugo{\l}\k{e}cki}}, \bibinfo
  {author} {\bibfnamefont {R.}~\bibnamefont {Gonz{\'a}lez-F{\'e}rez}},\ and\
  \bibinfo {author} {\bibfnamefont {J.}~\bibnamefont {K\"upper}},\ }\bibfield
  {title} {\enquote {\bibinfo {title} {Adiabatic mixed-field orientation of
  ground-state-selected carbonyl sulfide molecules},}\ }\href
  {https://doi.org/cphc.201600710R2} {\bibfield  {journal} {\bibinfo  {journal}
  {Chem. Phys. Chem.}\ }\textbf {\bibinfo {volume} {17}},\ \bibinfo {pages}
  {3740--3746} (\bibinfo {year} {2016})},\ \Eprint
  {https://arxiv.org/abs/1607.05615} {arXiv:1607.05615 [physics]} \BibitemShut
  {NoStop}%
\bibitem [{\citenamefont {Filsinger}\ \emph
  {et~al.}(2009{\natexlab{b}})\citenamefont {Filsinger}, \citenamefont
  {K{\"u}pper}, \citenamefont {Meijer}, \citenamefont {Holmegaard},
  \citenamefont {Nielsen}, \citenamefont {Nevo}, \citenamefont {Hansen},\ and\
  \citenamefont {Stapelfeldt}}]{Filsinger:JCP131:064309}%
  \BibitemOpen
  \bibfield  {author} {\bibinfo {author} {\bibfnamefont {F.}~\bibnamefont
  {Filsinger}}, \bibinfo {author} {\bibfnamefont {J.}~\bibnamefont
  {K{\"u}pper}}, \bibinfo {author} {\bibfnamefont {G.}~\bibnamefont {Meijer}},
  \bibinfo {author} {\bibfnamefont {L.}~\bibnamefont {Holmegaard}}, \bibinfo
  {author} {\bibfnamefont {J.~H.}\ \bibnamefont {Nielsen}}, \bibinfo {author}
  {\bibfnamefont {I.}~\bibnamefont {Nevo}}, \bibinfo {author} {\bibfnamefont
  {J.~L.}\ \bibnamefont {Hansen}},\ and\ \bibinfo {author} {\bibfnamefont
  {H.}~\bibnamefont {Stapelfeldt}},\ }\bibfield  {title} {\enquote {\bibinfo
  {title} {Quantum-state selection, alignment, and orientation of large
  molecules using static electric and laser fields},}\ }\href
  {https://doi.org/10.1063/1.3194287} {\bibfield  {journal} {\bibinfo
  {journal} {J. Chem. Phys.}\ }\textbf {\bibinfo {volume} {131}},\ \bibinfo
  {pages} {064309} (\bibinfo {year} {2009}{\natexlab{b}})},\ \Eprint
  {https://arxiv.org/abs/0903.5413} {arXiv:0903.5413 [physics]} \BibitemShut
  {NoStop}%
\bibitem [{\citenamefont {Hankin}\ \emph {et~al.}(2001)\citenamefont {Hankin},
  \citenamefont {Villeneuve}, \citenamefont {Corkum},\ and\ \citenamefont
  {Rayner}}]{Hankin:PRA64:013405}%
  \BibitemOpen
  \bibfield  {author} {\bibinfo {author} {\bibfnamefont {S.}~\bibnamefont
  {Hankin}}, \bibinfo {author} {\bibfnamefont {D.}~\bibnamefont {Villeneuve}},
  \bibinfo {author} {\bibfnamefont {P.}~\bibnamefont {Corkum}},\ and\ \bibinfo
  {author} {\bibfnamefont {D.}~\bibnamefont {Rayner}},\ }\bibfield  {title}
  {\enquote {\bibinfo {title} {Intense-field laser ionization rates in atoms
  and molecules},}\ }\href {https://doi.org/10.1103/PhysRevA.64.013405}
  {\bibfield  {journal} {\bibinfo  {journal} {Phys. Rev. A}\ }\textbf {\bibinfo
  {volume} {64}},\ \bibinfo {pages} {013405} (\bibinfo {year}
  {2001})}\BibitemShut {NoStop}%
\bibitem [{\citenamefont {Wiese}\ \emph {et~al.}(2019)\citenamefont {Wiese},
  \citenamefont {Olivieri}, \citenamefont {Trabattoni}, \citenamefont
  {Trippel},\ and\ \citenamefont {Küpper}}]{Wiese:NJP21:083011}%
  \BibitemOpen
  \bibfield  {author} {\bibinfo {author} {\bibfnamefont {J.}~\bibnamefont
  {Wiese}}, \bibinfo {author} {\bibfnamefont {J.-F.}\ \bibnamefont {Olivieri}},
  \bibinfo {author} {\bibfnamefont {A.}~\bibnamefont {Trabattoni}}, \bibinfo
  {author} {\bibfnamefont {S.}~\bibnamefont {Trippel}},\ and\ \bibinfo {author}
  {\bibfnamefont {J.}~\bibnamefont {Küpper}},\ }\bibfield  {title} {\enquote
  {\bibinfo {title} {Strong-field photoelectron momentum imaging of {OCS} at
  finely resolved incident intensities},}\ }\href
  {https://doi.org/10.1088/1367-2630/ab34e8} {\bibfield  {journal} {\bibinfo
  {journal} {New J. Phys.}\ }\textbf {\bibinfo {volume} {21}},\ \bibinfo
  {pages} {083011} (\bibinfo {year} {2019})},\ \Eprint
  {https://arxiv.org/abs/1904.07519} {arXiv:1904.07519 [physics]} \BibitemShut
  {NoStop}%
\bibitem [{\citenamefont {Fehre}\ \emph {et~al.}(2018)\citenamefont {Fehre},
  \citenamefont {Trojanowskaja}, \citenamefont {Gatzke}, \citenamefont
  {Kunitski}, \citenamefont {Trinter}, \citenamefont {Zeller}, \citenamefont
  {Schmidt}, \citenamefont {Stohner}, \citenamefont {Berger}, \citenamefont
  {Czasch} \emph {et~al.}}]{Fehre:RSI89:045112}%
  \BibitemOpen
  \bibfield  {author} {\bibinfo {author} {\bibfnamefont {K.}~\bibnamefont
  {Fehre}}, \bibinfo {author} {\bibfnamefont {D.}~\bibnamefont
  {Trojanowskaja}}, \bibinfo {author} {\bibfnamefont {J.}~\bibnamefont
  {Gatzke}}, \bibinfo {author} {\bibfnamefont {M.}~\bibnamefont {Kunitski}},
  \bibinfo {author} {\bibfnamefont {F.}~\bibnamefont {Trinter}}, \bibinfo
  {author} {\bibfnamefont {S.}~\bibnamefont {Zeller}}, \bibinfo {author}
  {\bibfnamefont {L.~P.~H.}\ \bibnamefont {Schmidt}}, \bibinfo {author}
  {\bibfnamefont {J.}~\bibnamefont {Stohner}}, \bibinfo {author} {\bibfnamefont
  {R.}~\bibnamefont {Berger}}, \bibinfo {author} {\bibfnamefont
  {A.}~\bibnamefont {Czasch}}, \emph {et~al.},\ }\bibfield  {title} {\enquote
  {\bibinfo {title} {Absolute ion detection efficiencies of microchannel plates
  and funnel microchannel plates for multi-coincidence detection},}\ }\href
  {https://doi.org/10.1063/1.5022564} {\bibfield  {journal} {\bibinfo
  {journal} {Rev. Sci. Instrum.}\ }\textbf {\bibinfo {volume} {89}},\ \bibinfo
  {pages} {045112} (\bibinfo {year} {2018})}\BibitemShut {NoStop}%
\bibitem [{\citenamefont {Rivero}, \citenamefont {Meuwly},\ and\ \citenamefont
  {Willitsch}(2017)}]{Rivero:CPL683:598}%
  \BibitemOpen
  \bibfield  {author} {\bibinfo {author} {\bibfnamefont {U.}~\bibnamefont
  {Rivero}}, \bibinfo {author} {\bibfnamefont {M.}~\bibnamefont {Meuwly}},\
  and\ \bibinfo {author} {\bibfnamefont {S.}~\bibnamefont {Willitsch}},\
  }\bibfield  {title} {\enquote {\bibinfo {title} {A computational study of the
  diels-alder reactions between 2,3-dibromo-1,3-butadiene and maleic
  anhydride},}\ }\href
  {https://doi.org/https://doi.org/10.1016/j.cplett.2017.03.063} {\bibfield
  {journal} {\bibinfo  {journal} {Chem. Phys. Lett.}\ }\textbf {\bibinfo
  {volume} {683}},\ \bibinfo {pages} {598 -- 605} (\bibinfo {year} {2017})},\
  \bibinfo {note} {ahmed Zewail (1946-2016) Commemoration Issue of Chemical
  Physics Letters}\BibitemShut {NoStop}%
\end{thebibliography}%


%aipnum4-2.bst 2019-01-14 (MD) hand-edited version of apsrev4-1.bst
\begin{thebibliography}{5}%
\makeatletter
\providecommand \@ifxundefined [1]{%
 \@ifx{#1\undefined}
}%
\providecommand \@ifnum [1]{%
 \ifnum #1\expandafter \@firstoftwo
 \else \expandafter \@secondoftwo
 \fi
}%
\providecommand \@ifx [1]{%
 \ifx #1\expandafter \@firstoftwo
 \else \expandafter \@secondoftwo
 \fi
}%
\providecommand \natexlab [1]{#1}%
\providecommand \enquote  [1]{``#1''}%
\providecommand \bibnamefont  [1]{#1}%
\providecommand \bibfnamefont [1]{#1}%
\providecommand \citenamefont [1]{#1}%
\providecommand \href@noop [0]{\@secondoftwo}%
\providecommand \href [0]{\begingroup \@sanitize@url \@href}%
\providecommand \@href[1]{\@@startlink{#1}\@@href}%
\providecommand \@@href[1]{\endgroup#1\@@endlink}%
\providecommand \@sanitize@url [0]{\catcode `\\12\catcode `\$12\catcode
  `\&12\catcode `\#12\catcode `\^12\catcode `\_12\catcode `\%12\relax}%
\providecommand \@@startlink[1]{}%
\providecommand \@@endlink[0]{}%
\providecommand \url  [0]{\begingroup\@sanitize@url \@url }%
\providecommand \@url [1]{\endgroup\@href {#1}{\urlprefix }}%
\providecommand \urlprefix  [0]{URL }%
\providecommand \Eprint [0]{\href }%
\providecommand \doibase [0]{https://doi.org/}%
\providecommand \selectlanguage [0]{\@gobble}%
\providecommand \bibinfo  [0]{\@secondoftwo}%
\providecommand \bibfield  [0]{\@secondoftwo}%
\providecommand \translation [1]{[#1]}%
\providecommand \BibitemOpen [0]{}%
\providecommand \bibitemStop [0]{}%
\providecommand \bibitemNoStop [0]{.\EOS\space}%
\providecommand \EOS [0]{\spacefactor3000\relax}%
\providecommand \BibitemShut  [1]{\csname bibitem#1\endcsname}%
\let\auto@bib@innerbib\@empty
%</preamble>
\bibitem [{\citenamefont {Chang}\ \emph {et~al.}(2014)\citenamefont {Chang},
  \citenamefont {Filsinger}, \citenamefont {Sartakov}, and\ \citenamefont
  {K{\"u}pper}}]{Chang:CPC185:339}%
  \BibitemOpen
  \bibfield  {author} {\bibinfo {author} {\bibfnamefont {Y.-P.}\ \bibnamefont
  {Chang}}, \bibinfo {author} {\bibfnamefont {F.}~\bibnamefont {Filsinger}},
  \bibinfo {author} {\bibfnamefont {B.}~\bibnamefont {Sartakov}}, and\
  \bibinfo {author} {\bibfnamefont {J.}~\bibnamefont {K{\"u}pper}},\ }\bibfield
   {title} {\enquote {\bibinfo {title} {\textsc{CMIstark}: {P}ython package for
  the {S}tark-effect calculation and symmetry classification of linear,
  symmetric and asymmetric top wavefunctions in dc electric fields},}\ }\href
  {https://doi.org/10.1016/j.cpc.2013.09.001} {\bibfield  {journal} {\bibinfo
  {journal} {Comp. Phys. Comm.}\ }\textbf {\bibinfo {volume} {185}},\ \bibinfo
  {pages} {339--349} (\bibinfo {year} {2014})},\ \bibinfo {note} {current
  version available from \href{https://github.com/CFEL-CMI/cmistark}{GitHub}},\
  \Eprint {https://arxiv.org/abs/1308.4076} {arXiv:1308.4076 [physics]}\BibitemShut {NoStop}%
\bibitem [{\citenamefont {Foster}, \citenamefont {Rao}, and\ \citenamefont
  {Curl~Jr}(1965)}]{Foster:JCP43:1064}%
  \BibitemOpen
  \bibfield  {author} {\bibinfo {author} {\bibfnamefont {P.~D.}\ \bibnamefont
  {Foster}}, \bibinfo {author} {\bibfnamefont {V.~M.}\ \bibnamefont {Rao}},\
  and\ \bibinfo {author} {\bibfnamefont {R.~F.}\ \bibnamefont {Curl~Jr}},\
  }\bibfield  {title} {\enquote {\bibinfo {title} {Microwave spectrum of methyl
  vinyl ketone},}\ }\href {https://doi.org/10.1063/1.1696821} {\bibfield
  {journal} {\bibinfo  {journal} {J. Chem. Phys.}\ }\textbf {\bibinfo {volume}
  {43}},\ \bibinfo {pages} {1064--1066} (\bibinfo {year} {1965})}\BibitemShut
  {NoStop}%
\bibitem [{\citenamefont {Wilcox}\ \emph {et~al.}(2011)\citenamefont {Wilcox},
  \citenamefont {Shirar}, \citenamefont {Williams}, and\ \citenamefont
  {Dian}}]{Wilcox:CPL508:10}%
  \BibitemOpen
  \bibfield  {author} {\bibinfo {author} {\bibfnamefont {D.~S.}\ \bibnamefont
  {Wilcox}}, \bibinfo {author} {\bibfnamefont {A.~J.}\ \bibnamefont {Shirar}},
  \bibinfo {author} {\bibfnamefont {O.~L.}\ \bibnamefont {Williams}}, and\
  \bibinfo {author} {\bibfnamefont {B.~C.}\ \bibnamefont {Dian}},\ }\bibfield
  {title} {\enquote {\bibinfo {title} {Additional conformer observed in the
  microwave spectrum of methyl vinyl ketone},}\ }\href
  {https://doi.org/10.1016/j.cplett.2011.04.001} {\bibfield  {journal}
  {\bibinfo  {journal} {Chem. Phys. Lett.}\ }\textbf {\bibinfo {volume}
  {508}},\ \bibinfo {pages} {10--16} (\bibinfo {year} {2011})}\BibitemShut
  {NoStop}%
\bibitem [{\citenamefont {Filsinger}\ \emph {et~al.}(2009)\citenamefont
  {Filsinger}, \citenamefont {K{\"u}pper}, \citenamefont {Meijer},
  \citenamefont {Holmegaard}, \citenamefont {Nielsen}, \citenamefont {Nevo},
  \citenamefont {Hansen}, and\ \citenamefont
  {Stapelfeldt}}]{Filsinger:JCP131:064309}%
  \BibitemOpen
  \bibfield  {author} {\bibinfo {author} {\bibfnamefont {F.}~\bibnamefont
  {Filsinger}}, \bibinfo {author} {\bibfnamefont {J.}~\bibnamefont
  {K{\"u}pper}}, \bibinfo {author} {\bibfnamefont {G.}~\bibnamefont {Meijer}},
  \bibinfo {author} {\bibfnamefont {L.}~\bibnamefont {Holmegaard}}, \bibinfo
  {author} {\bibfnamefont {J.~H.}\ \bibnamefont {Nielsen}}, \bibinfo {author}
  {\bibfnamefont {I.}~\bibnamefont {Nevo}}, \bibinfo {author} {\bibfnamefont
  {J.~L.}\ \bibnamefont {Hansen}}, and\ \bibinfo {author} {\bibfnamefont
  {H.}~\bibnamefont {Stapelfeldt}},\ }\bibfield  {title} {\enquote {\bibinfo
  {title} {Quantum-state selection, alignment, and orientation of large
  molecules using static electric and laser fields},}\ }\href
  {https://doi.org/10.1063/1.3194287} {\bibfield  {journal} {\bibinfo
  {journal} {J. Chem. Phys.}\ }\textbf {\bibinfo {volume} {131}},\ \bibinfo
  {pages} {064309} (\bibinfo {year} {2009})},\ \Eprint
  {https://arxiv.org/abs/0903.5413} {arXiv:0903.5413 [physics]}\BibitemShut
  {NoStop}%
\bibitem [{\citenamefont {Chang}\ \emph {et~al.}(2015)\citenamefont {Chang},
  \citenamefont {Horke}, \citenamefont {Trippel}, and\ \citenamefont
  {Küpper}}]{Chang:IRPC34:557}%
  \BibitemOpen
  \bibfield  {author} {\bibinfo {author} {\bibfnamefont {Y.-P.}\ \bibnamefont
  {Chang}}, \bibinfo {author} {\bibfnamefont {D.~A.}\ \bibnamefont {Horke}},
  \bibinfo {author} {\bibfnamefont {S.}~\bibnamefont {Trippel}}, and\ \bibinfo
  {author} {\bibfnamefont {J.}~\bibnamefont {Küpper}},\ }\bibfield  {title}
  {\enquote {\bibinfo {title} {Spatially-controlled complex molecules and their
  applications},}\ }\href {https://doi.org/10.1080/0144235X.2015.1077838}
  {\bibfield  {journal} {\bibinfo  {journal} {Int. Rev. Phys. Chem.}\ }\textbf
  {\bibinfo {volume} {34}},\ \bibinfo {pages} {557--590} (\bibinfo {year}
  {2015})},\ \Eprint {https://arxiv.org/abs/1505.05632} {arXiv:1505.05632
  [physics]}\BibitemShut {NoStop}%
\end{thebibliography}%


%aipnum4-2.bst 2019-01-14 (MD) hand-edited version of apsrev4-1.bst
\end{document}

\title{Supplemental Material: Spatial Separation of the Conformers of Methyl Vinyl Ketone}%
\author{Jia Wang}\tsinghua\cfeldesy%
\author{Ardita Kilaj}\basel%
\author{Lanhai He}\jilinu\cfeldesy%
\author{Karol Długołęcki}\cfeldesy%
\author{Stefan Willitsch}\swemail\basel%
\author{Jochen Küpper}\jkemail\cmiweb\cfeldesy\uhhphys\uhhcui%
%\date{\today}%
\maketitle
\enlargethispage{2mm}

\section{Sample preparation}
\begin{figure}
   \includegraphics[width=\linewidth]{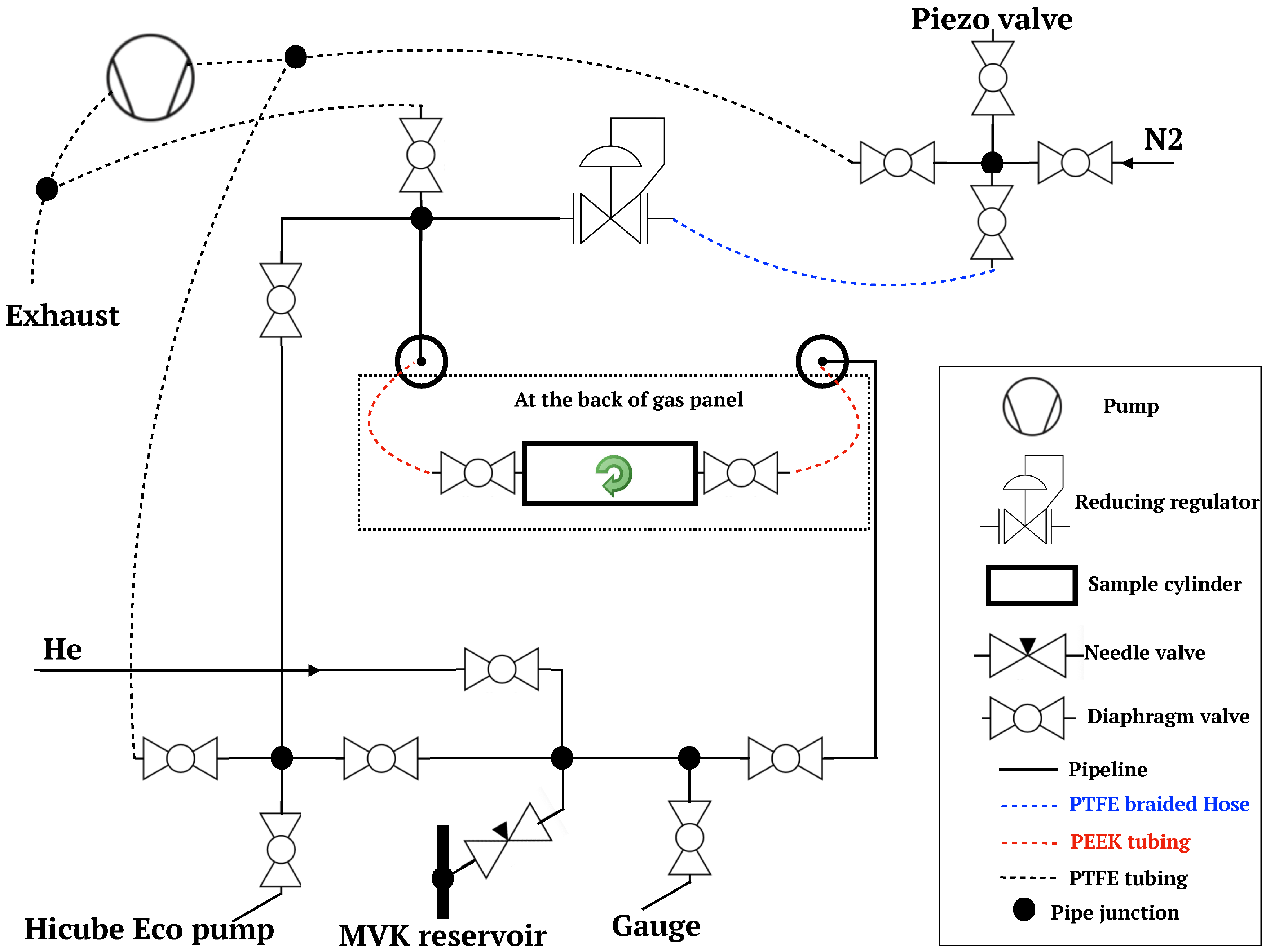}
   \caption{Schematic of the gas handling system, see text for details.}
   \label{fig:gaspaneldesign}
\end{figure}

MVK is a liquid with a relatively high vapor pressure of $\ordsim130$~mbar at room temperature and
its vapor/air mixtures are explosive. \autoref{fig:gaspaneldesign} shows the schematic of the gas
panel for sample preparation. This includes the pumping system, pipes, the MVK reservoir, and a
rotating sample cylinder. The sample cylinder is mounted on a rotating motor (DGM130R-AZAC, Oriental
Motor) and is connected by soft PEEK tubes. The 2~ml MVK sample is filled into the MVK reservoir and
de-aired. First, the whole gas line is evacuated (HiCube Eco, Pfeiffer Vacuum) to
$\sim10^{-2}$~mbar, then MVK vapor is leaked into the sample cylinder to the designed pressure, and
subsequently filled with helium to 20~bar. The cylinder is closed and rotated. To minimize corrosion
the MVK sample is removed from the gas lines and the sample reservoir when not used.

\begin{figure}[b]
   \includegraphics[width=\linewidth]{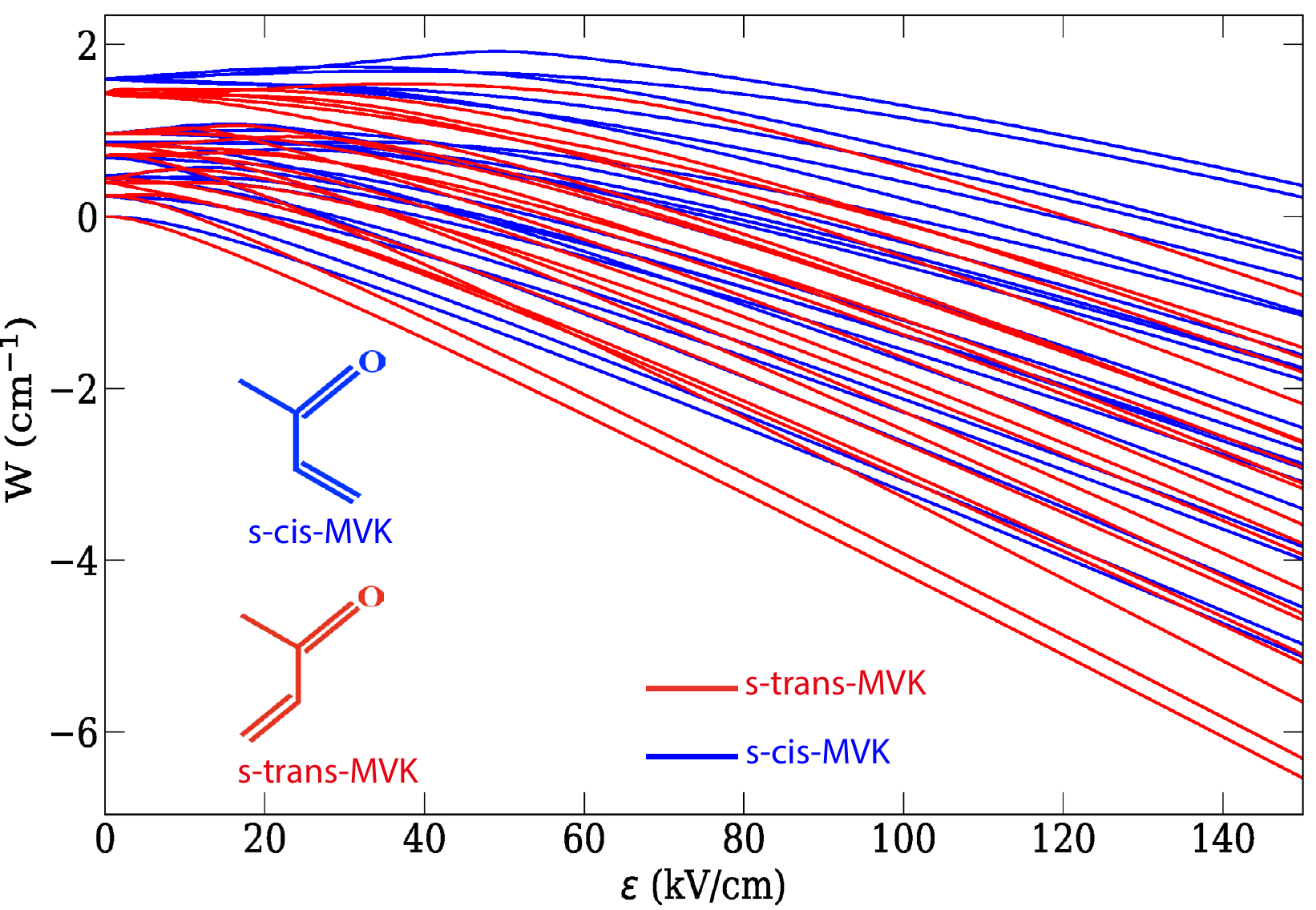}%
   \caption{The calculated adiabatic Stark energies of the lowest-energy rotational states
      ($J=0\ldots2$) of MVK in a dc electric field.}
   \label{fig:starkenergy}
\end{figure}

\begin{figure*}
   \includegraphics[width=\linewidth]{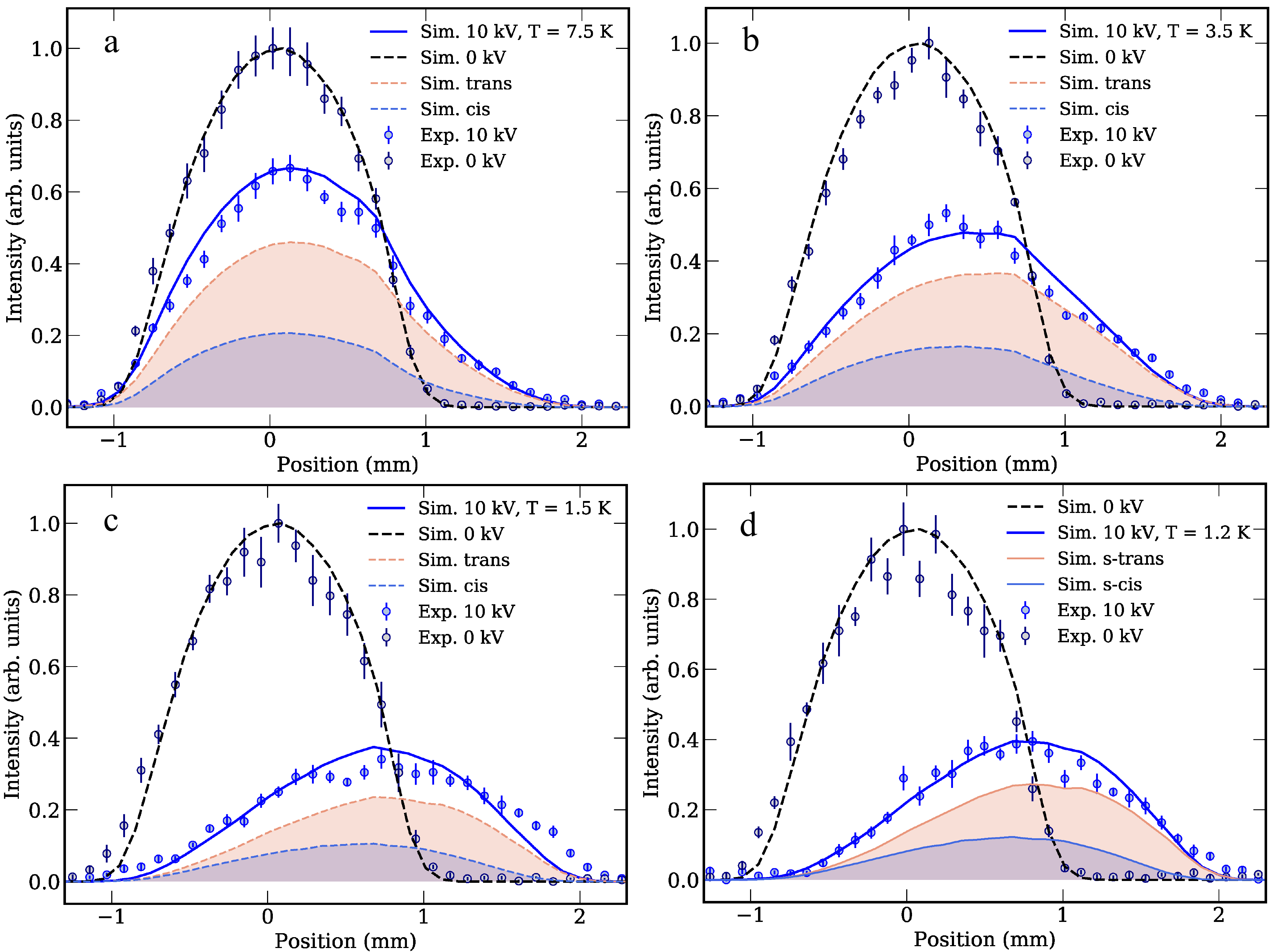}%
   \caption{Experimental and simulated deflection profiles of MVK seeded in helium at different
      pressures of (a) 2~bar, (b) 4~bar, (c) 6~bar, and (d) 8~bar.}
   \label{fig:deflectioncom}
\end{figure*}

\section{Stark energy calculations and numerical simulations}
The Stark energies for all rotational states of both conformers of MVK up to $J=15$ were calculated
on a basis of all field-free rotational states with $J\leq30$ using the freely available
\textsc{CMIstark} software package~\cite{Chang:CPC185:339}. The rotational constants and dipole
moments of \cisMVK and \transMVK~\cite{Foster:JCP43:1064, Wilcox:CPL508:10} are summarized in
\autoref{tab:constants}. The Stark curves of the lowest-energy rotational states are shown in
\autoref{fig:starkenergy}.

\autoref{fig:deflectioncom} shows the normalized experimental and simulated vertical molecular beam
profiles of MVK seeded in 2, 4, 6, and 8~bar of helium, for voltages to the deflector of 0~V and
10~kV. The mixing ratio MVK/helium of the sample was 200~ppm. Solid and dotted lines in
\autoref{fig:deflectioncom} show simulated spatial profiles of the individual conformers. We
simulated 1 $\times10^{5}$ classical trajectories for every quantum state with $J\leq14$ and the
experimental parameters~\cite{Filsinger:JCP131:064309, Chang:IRPC34:557}. Assuming a Boltzmann
population distribution of rotational states, the rotational temperatures of MVK seeded in helium of
different pressures (2, 4, 6, 8~bar) were fitted to be 7.5, 3.5, 1.5, and 1.2~K.
\begin{table}[b]
   \centering%
   \caption{Rotational constants and dipole moment components~\cite{Foster:JCP43:1064,
         Wilcox:CPL508:10} used for the Stark effect calculations.}
   \begin{tabular}{l@{\hspace{1em}}c@{\hspace{1em}}c}
     \hline\hline
     & \textit{s-cis} conformer  & \textit{s-trans} conformer \\
     \hline
     A (MHz)             & 10240.938    & 8941.590        \\
     B (MHz)             & ~~~3991.6351    & ~~4274.5443        \\
     C (MHz)             & ~~2925.648    & ~~2945.3315        \\
     $\mu_\text{A}$ (D)  & -0.57         & ~2.53            \\
     $\mu_\text{B}$ (D)  & ~2.88         & -1.91           \\
     $\mu_\text{C}$ (D)  & 0         & 0 \\
     \hline\hline
   \end{tabular}
   \label{tab:constants}
\end{table}

\bibliography{string,cmi}

\clearpage

\begin{figure}
   \includegraphics[width=\linewidth]{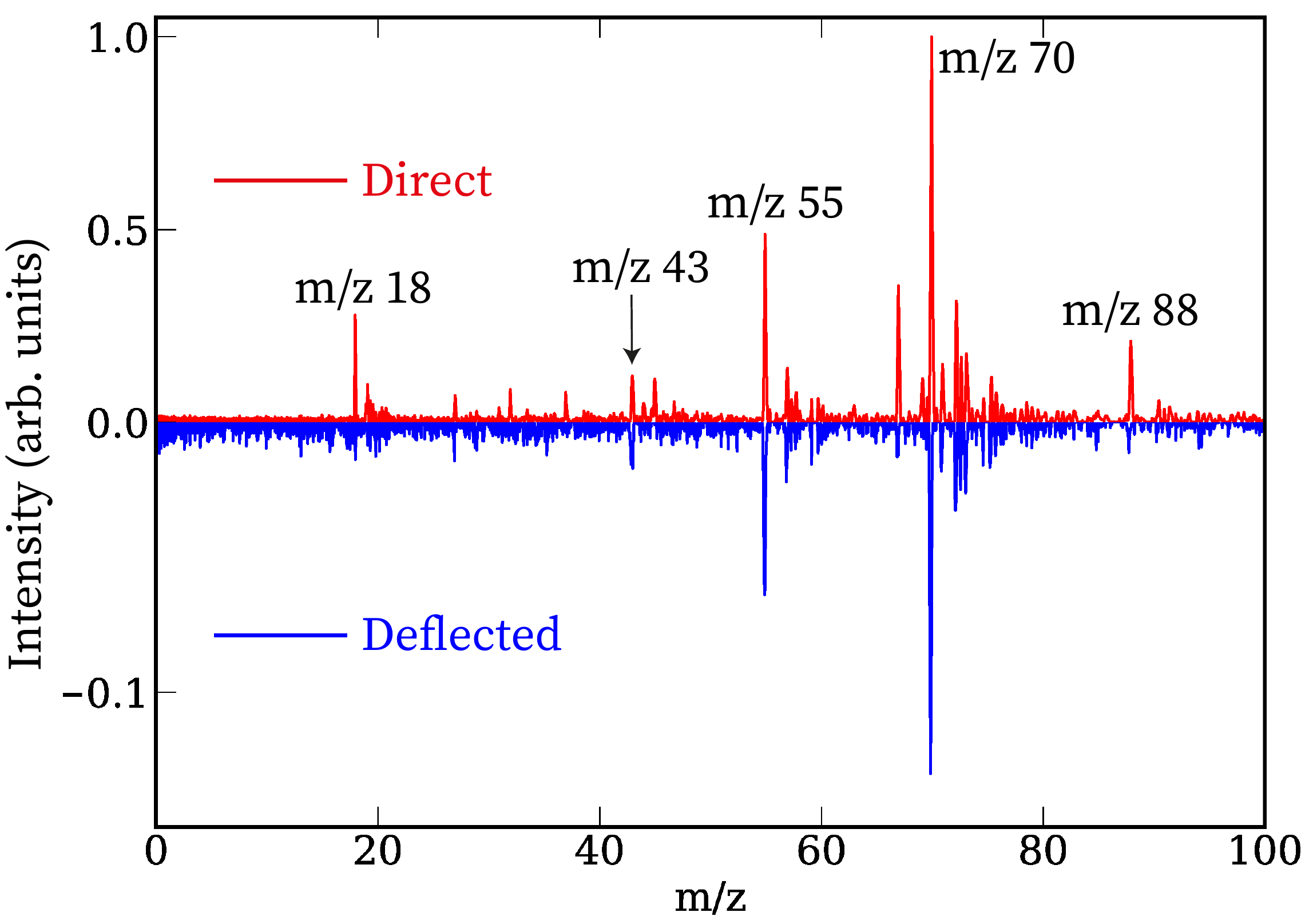}%
   \caption{Mass spectrum of the direct (0~kV) and deflected (10~kV) molecular beam, corresponding
      to the positions at 0~mm and 1.65~mm in Figure~2 in the main manuscript. The spectra were
      normalized to the monomer ion signal ($m/z=70$). No ion signal was found above $m/z=100$.}
   \label{fig:MS}
\end{figure}

\begin{figure}
   \includegraphics[width=\linewidth]{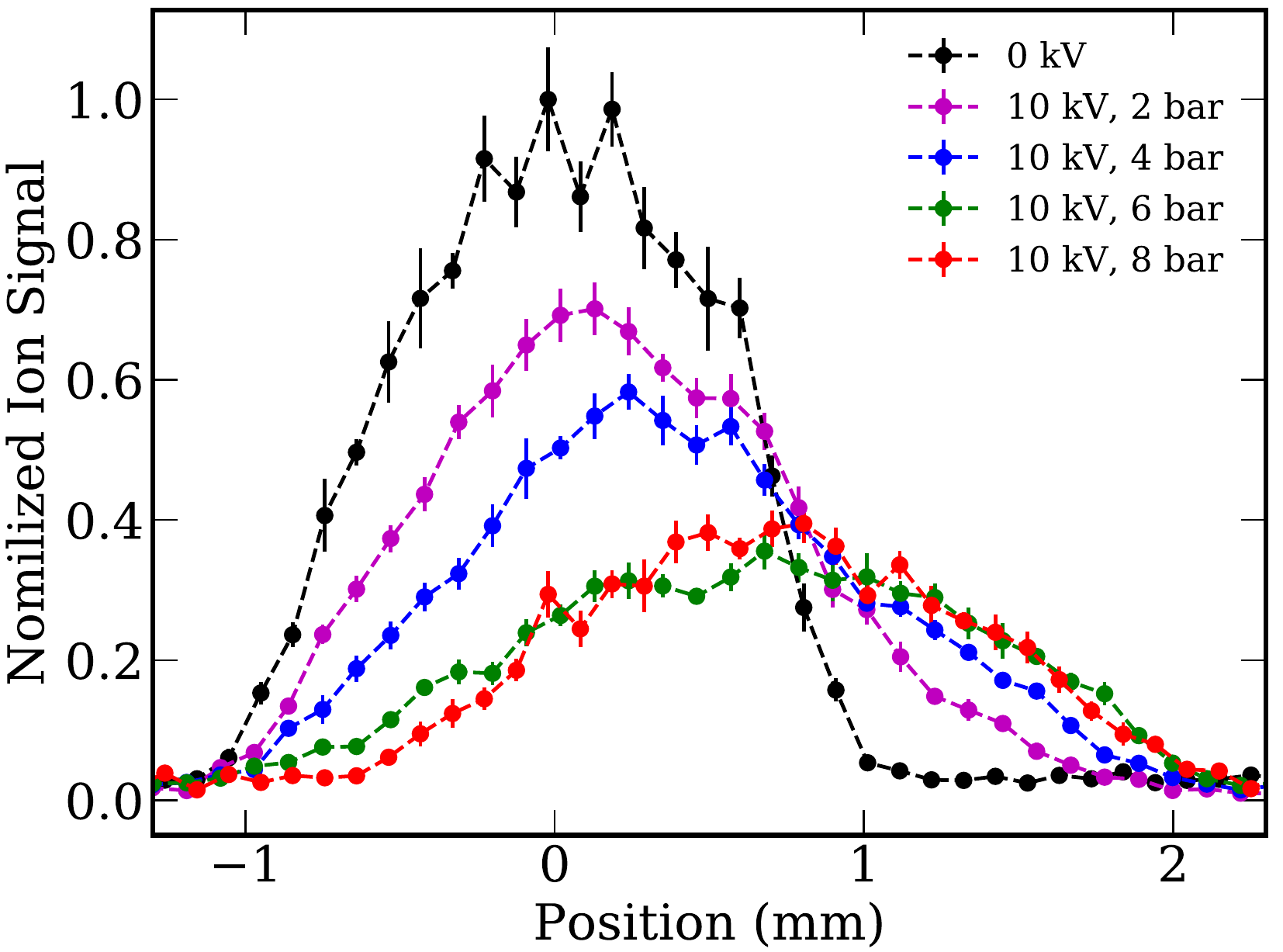}%
   \caption{Vertical profiles of the direct (0~kV) and deflected (10~kV) molecular beam of MVK for
      different backing pressures applied to the piezo valve.}
   \label{fig:pressure}
\end{figure}